\documentclass{jfm}
\usepackage{color}
\usepackage{graphicx}
\usepackage{natbib}
\usepackage{amsmath}
\usepackage{caption}
\usepackage{subcaption}

\usepackage{booktabs}
\usepackage[table,xcdraw]{xcolor}
\usepackage{booktabs, multirow} 
\usepackage{rotating,tabularx}
\usepackage{multirow}
\usepackage{changepage,threeparttable}
\usepackage{makecell}
\usepackage{soul}

\title{Buoyancy driven bubbly flows: role of meso-scale structures on the relative motion between phases in bubble columns operated in the heterogeneous regime}

\author[Y. Mezui, M. Obligado and A. Cartellier]%
       {Y. Mezui\aff{1}, M. Obligado\aff{1} and A. Cartellier\aff{1}
         \corresp{Email address for correspondence:  alain.cartellier@cnrs.fr}}

\affiliation{\aff{1} Universit\'{e} Grenoble Alpes, CNRS, Grenoble-INP, LEGI, F-38000, Grenoble, France }

\pubyear{2010}
\volume{650}
\pagerange{119--126}
\date{?; revised ?; accepted ?. - To be entered by editorial office}

\begin{document}
\sloppy 

\maketitle
\begin{abstract}

The hydrodynamics of bubble columns in the heterogeneous regime is investigated from experiments with bubbles at large particle Reynolds numbers and without coalescence. The void fraction field $\varepsilon$ at small scales, analyzed with Vorono\"i tessellations, corresponds to a Random Poisson Process (RPP) in homogeneous conditions but it significantly differs from a RPP in the heterogeneous regime. The distance to a RPP allows identifying meso-scale structures, namely clusters, void regions and intermediate regions. A series of arguments demonstrate that the bubble motion is driven by the dynamics of these structures. Notably, bubbles in clusters (respectively in intermediate regions) are moving up faster, up to 3.5 (respectively 2) times the terminal velocity, than bubbles in void regions those absolute velocity equals the mean liquid velocity. Besides, the mean unconditional relative velocity of bubbles is recovered from mean relative velocities conditional to meso-scale structures, weighted by the proportion of bubbles in each structure. Assuming buoyancy-inertia equilibrium for each structure, the relative velocity is related with the characteristic size and concentration of meso-scale structures. By taking the latter quantities values at large gas superficial velocities, a cartoon of the internal flow structure is proposed. Arguments are put forward to help understanding why the relative velocity scales as $(gD\varepsilon)^{1/2}$ (with $D$ the column's diameter and $g$ gravity's acceleration). The proposed cartoon seems consistent with a fast-track mechanism that, for the moderate Rouse numbers studied, leads to liquid velocity fluctuations proportional to the relative velocity. The potential impact of coalescence on the above analysis is also commented.

\end{abstract}


\section{Introduction} \label{intro}

In bubble columns, gas is injected at the bottom of a vertical cylinder filled with liquid. Such systems are commonly employed as reactors (for chemical or biochemical transformations), mixers (metallurgy), separators (flotation), among several other applications. For low enough gas flow rate, bubbles are uniformly distributed over the column cross-section and gently rise up to the free surface. A characteristic of this so-called homogeneous regime is the linear increase of the void fraction $\varepsilon$ with the gas superfical velocity $V_{sg}$ (the latter is defined as the injected gas flow rate divided by the column cross-section). The apparent rise velocity, evaluated as $V_{sg}/\varepsilon$, is nearly constant in this regime, and its magnitude is of the order of the bubbles terminal velocity $U_T$. Increasing the inlet gas flow above some threshold, the flow becomes non-uniform, a mean recirculation settles at the reactor scale, and unsteady flow structures appear everywhere in the column, as noted by No\"el de Nevers in 1968~(\cite{de1968bubble}). In this so-called heterogeneous regime, the increase of the void fraction with $V_{sg}$ is significantly slowed down while the apparent rise velocity $V_{sg}/\varepsilon$ steadily increases with $V_{sg}$ and becomes much larger than $U_T$ (\cite{krishna1991model,ruzicka2013stability}). This heterogeneous regime is exploited in many applications, but its hydrodynamics remains poorly understood. In particular, it is of practical importance to know how to scale-up bubble columns from laboratory prototypes to actual industrial plants. But, and as shown by the successive reviews, notably from \cite{joshi1998gas,kantarci2005bubble,rollbusch_bubble_2015,kikukawa_physical_2017,besagni_two-phase_2018}, there is still no consensus on appropriate scaling rules.  

Recently, we have shown that, in the heterogeneous regime, buoyancy equilibrates inertia (\cite{Mezui2022}), and that feature leads to velocities scaling as $(gD\varepsilon)^{1/2}$, where $D$ is the bubble column diameter, $\varepsilon$ the void fraction and $g$ the gravitational acceleration. This scaling was shown to hold for mean velocities and for standard deviations. Besides, it applies to the liquid phase as well as to the gas phase. That result was found valid over a wide range of flow conditions (namely $0.1$m $\leq D \leq 3$m and $4-5$cm/s $\leq V_{sg} \leq 60$cm/s) when considering air-water systems involving bubbles with an equivalent diameter between about 3mm and 10mm (\cite{Mezui2022}). To complement that proposal, and inspired by a Zuber \& Findlay approach (\cite{zuber1965average}), the void fraction on the column axis $\varepsilon_{axis}$ was correlated with the Froude number (defined as $Fr=V_{sg}/(gD)^{1/2}$). A direct consequence of these findings, is that the relative velocity $U_R$, given as the difference between the mean gas and the mean liquid vertical velocities, is also evolving as $(gD\varepsilon)^{1/2}$: the relative velocity is thus expected to increase both with the bubble column diameter and with the void fraction. 

The change of the relative velocity with the void fraction is in qualitative agreement with experimentally observed trends since the apparent relative velocity either estimated as $V_{sg}/\varepsilon$  (\cite{krishna1991model,ruzicka2013stability}) or derived from a 1D kinematic approach (\cite{raimundo2019hydrodynamics}) monotonously increases with the gas superficial velocity and hence with void fraction. Besides, in Euler-Euler numerical simulations, the enhanced relative motion at large gas content is commonly enforced by introducing an \textit{ad-hoc} swarm coefficient (e.g. \cite{mcclure2017experimental,gemello2018cfd}) that quantifies the decrease of the drag force acting on a bubble with the local void fraction (\cite{ishii1979drag,simonnet2007experimental}). 

Concerning the impact of the bubble column diameter, we have no undisputable experimental evidence of the dependancy of $U_R$ on $D$. Moreover, to our knowledge, the swarm coefficients introduced in simulations never depend on $D$. Overall, it is not physically clear why $U_R$ should increase with the column diameter. 

Hereafter, we investigate the origin of the relative velocity in the heterogeneous regime. Some preliminary results presented by Maximiano Raimundo et al. (\cite{raimundo2019hydrodynamics}) indicate that concentration gradients should play a role in the flow dynamics. In particular, and as in turbulent convection, strong velocity differences were anticipated between zones in the flow with high void fraction and regions with low void fraction. To pursue the analysis, we need to get access to refined variables such as local concentration statistics and also to statistics on bubble velocity conditioned by the local concentration. To this end, we took benefit of the newly developed Doppler optical probe (a technology patented by A2 Photonic Sensors company) that simultaneously provides the gas phase indicator function and the translation velocity of bubbles (\cite{lefebvre2022new}). 

In section \ref{sec2}, the experimental conditions are presented and key variables such as void fraction and mean velocities characterising the bubble column behavior are provided. In section  \ref{sec3}, local void fraction measurements are introduced that pave the way to gather statistics on gas velocity measurements conditioned by the local gas concentration. Meso-scale structures are also presented together with their main characteristics. Section \ref{sec4} provides conditional bubble velocity measurements for clusters, intermediate regions and void regions and the contributions of these meso-scale structures to the absolute bubble velocity and to the relative bubble velocity are discussed. In section \ref{sec5}, a model relating bubble conditional velocities with the size and the concentration of the corresponding meso-scale structures is proposed and successfully tested. The relevance of a $(gD\varepsilon)^{1/2}$ scaling for the relative velocity is also debated. Finally, in Section  \ref{sec6}, we show that, according to the internal flow topology, a fast-track mechanism is potentially at play that would explain why velocity fluctuations also scale as $(gD\varepsilon)^{1/2}$ in heterogeneous bubble columns.

\section{Experimental conditions and unconditional velocity measurements \label{sec2}}

The experiment is the one exploited in \cite{Mezui2022}. It consists in a $3$m high and $D=0.4$m internal diameter air-water bubble column. The water quality was such that coalescence was absent or at least very weak. The injector is a 10mm thick plexiglass plate perforated by $352$ orifices (with 1mm internal diameter) uniformly distributed over the cross-section. The static liquid height $H_0$ was set to $2.02$m, a value large enough to avoid any sensitivity of measurements to $H_0$. Experiments were performed for $V_{sg}$ ranging from 0.6cm/s to 26cm/s. Information relative to bubbles were acquired with a Doppler probe (\cite{lefebvre2022new}). Such probe ensures the detection of phases with a high resolution (its latency length is $\approx 6~\mu m$) and it provides the velocity of bubbles. Liquid phase statistics were obtained from a Pavlov tube (see \cite{Mezui2022} for additional details). Over the range of flow conditions considered, the mean equivalent bubble diameter remained within the interval $\left[ 6.62mm; 7.35mm \right]$, so that the particle Reynolds number evolved in the range $1450-1550$. Measurements were achieved in the quasi fully developed region at $H/D=3.625$, where transverse profiles of velocity and void fraction are self-similar when normalized by the relevant value taken on the column axis. 

\begin{figure}
\centering
\includegraphics[width=\textwidth]{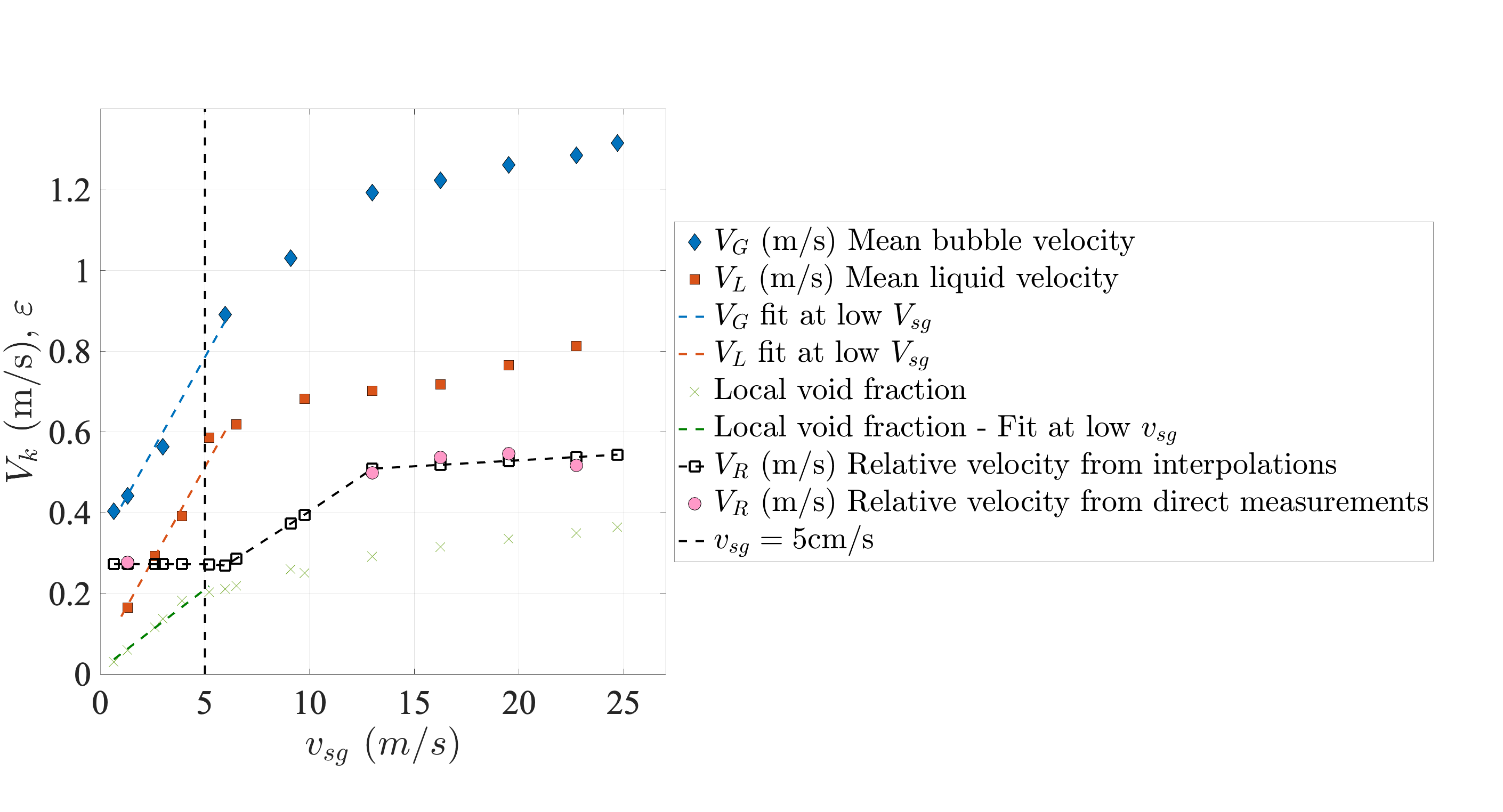}
\caption{Evolution of the void fraction, and of mean vertical velocities of bubbles $V_G$ and of the liquid $V_L$ with the gas superficial velocity $V_{sg}$. The relative velocity has been derived from raw measurements performed at the same $V_{sg}$ (pink dots), and from interpolations of phasic velocities versus $V_{sg}$ (black squares). The straight lines in the homogeneous regime are linear fits of the data. Measurements performed in a $D=0.4$m column, at $H/D=3.625$ and on the column axis. The vertical black dashed line indicates the transition from the homogeneous to the heterogeneous regime at $v_{sg} \sim 5cm/s$.} \label{fig:fig1}
\end{figure}

To qualify the bubble column behaviour, we report in Figure \ref{fig:fig1} the void fraction and the mean bubble and liquid vertical velocities on the axis of the column versus the superficial gas velocity $v_{sg}$. The mean velocities correspond to statistics combining upward and downward motions (see the discussion in \cite{Mezui2022}). The homogeneous-heterogeneous transition is indicated by the vertical dash line at $V_{sg}=5$cm/s. Direct relative velocity measurements gathered whenever $V_G$ and $V_L$ data were available for the same gas superficial velocity are also plotted in Figure \ref{fig:fig1} (see pink dots). We also use interpolations of $U_G$ and $V_L$ to estimate the relative velocity for others $V_{sg}$ values. In particular, the linear fits of mean bubble and liquid velocities in the homogeneous regime are parallel indicating that the relative velocity is constant in that regime. The latter amounts to $\approx 27$cm/s, which is close to the terminal velocity $U_T$. From the transition, the relative velocity neatly increases with $V_{sg}$, up to $V_{sg}\approx 13$cm/s. At larger $V_{sg}$, that is deeper in the heterogeneous regime, the relative velocity happens to nearly stabilize at about $2.3-2.5 U_T$. Note, however, that in that range ,and according to the few available data, the relative velocity is still smoothly increasing with $V_{sg}$. 

Overall, Figure \ref{fig:fig1} indisputably demonstrates that the relative velocity increases well beyond $U_T$ in a bubble column operated in the heterogeneous regime. In that regime, and in terms of the scaling rule proposed by \cite{Mezui2022}, the data gathered in the $D=0.4$m column correspond to $V_G \approx1.09 (gD\varepsilon)^{1/2}$ and $V_L\approx 0.67 (gD\varepsilon)^{1/2}$ where $\varepsilon$ is the local void fraction on the column axis at $H/D=3.625$. The prefactors given here are derived from the data collected for $v_{sg} \geq 13$~cm/s, but, as shown in \cite{Mezui2022}, they hold over a significant range of column diameters and of flow conditions. Hence, the relative velocity in the heterogeneous regime and far enough from the transition is expected to behave as,

\begin{equation}\label{eq1N}
U_R \approx 0.41 (gD\varepsilon)^{1/2}.
\end{equation}

\noindent Equation \ref{eq1N} predicts that the relative velocity depends on the column diameter, a feature that is not trivial. In order to understand the origin of the relative velocity in these buoyancy driven bubbly flows, we focus our analysis on the connection between local concentration and bubble velocity. In the next section, a local void fraction is defined, and related statistics are discussed.

\section{Local void fraction and meso-scale structures \label{sec3}}


\subsection{Local void fraction and identification of meso-scale structures}

\begin{figure}
\centering
\includegraphics[width=\textwidth]{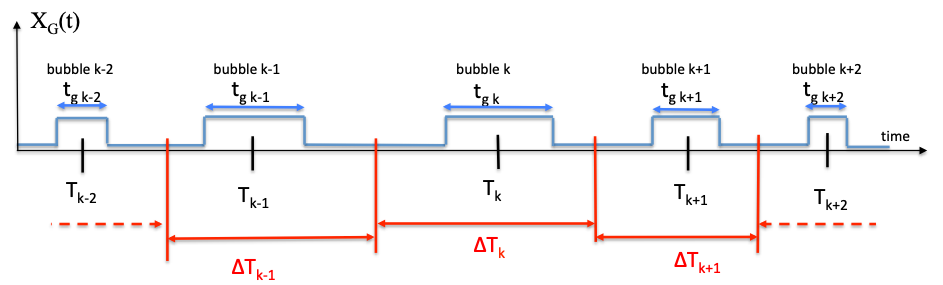}
\caption{Construction of a 1D Vorono\"i tessellation from the gas phase indicator function.} \label{fig14}
\end{figure}

Paralleling what we did for turbulent laden flows (\cite{monchaux2010preferential,sumbekova2017preferential,mora2018pitfalls}), we exploit 1D Vorono\"i tessellations built from the gas phase indicator function $X_G(t)$ (\cite{raimundo2015analyse,mezui2018characterization,raimundo2019hydrodynamics}). $X_G(t)$ is deduced from the signal delivered by an optical probe. For the gas phase indicator function measurements presented here, the probe orientation was held fixed (the probe was directed downwards). As shown in Figure \ref{fig14}, Vorono\"i cells are then built as successive time intervals, each containing a single bubble. For that, the centers $T_k$ of successive gas residence times $t_{gk}$ are identified. The mid-distance between successive centers $T_k$ and $T_{k+1}$ defines a Vorono\"i cell boundary. That process is repeated for all detected bubbles, and the width of the $k^{th}$ Vorono\"i cell that contains the $k^{th}$ bubble is given by $\Delta T_k = (T_{k+1} - T_{k-1})/2$. 

\begin{figure}
     \centering
     \begin{subfigure}[b]{0.45\textwidth}
         \centering
         \includegraphics[width=\textwidth]{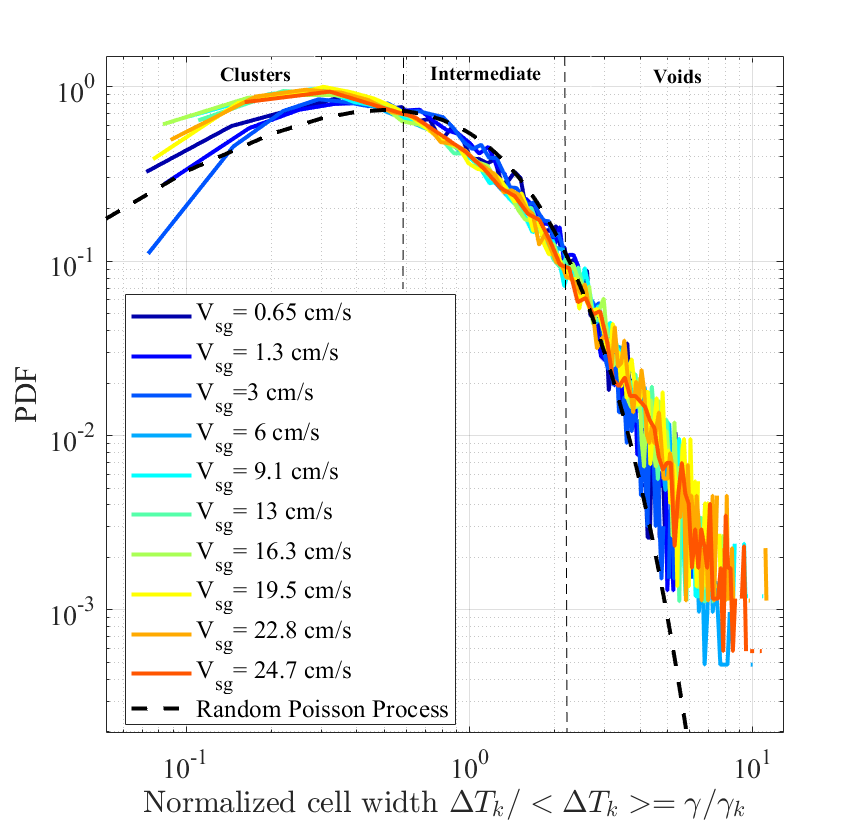}
         \caption{}
         \label{fig15a}
     \end{subfigure}
     \hfill
     \begin{subfigure}[b]{0.47\textwidth}
         \centering
         \includegraphics[width=\textwidth]{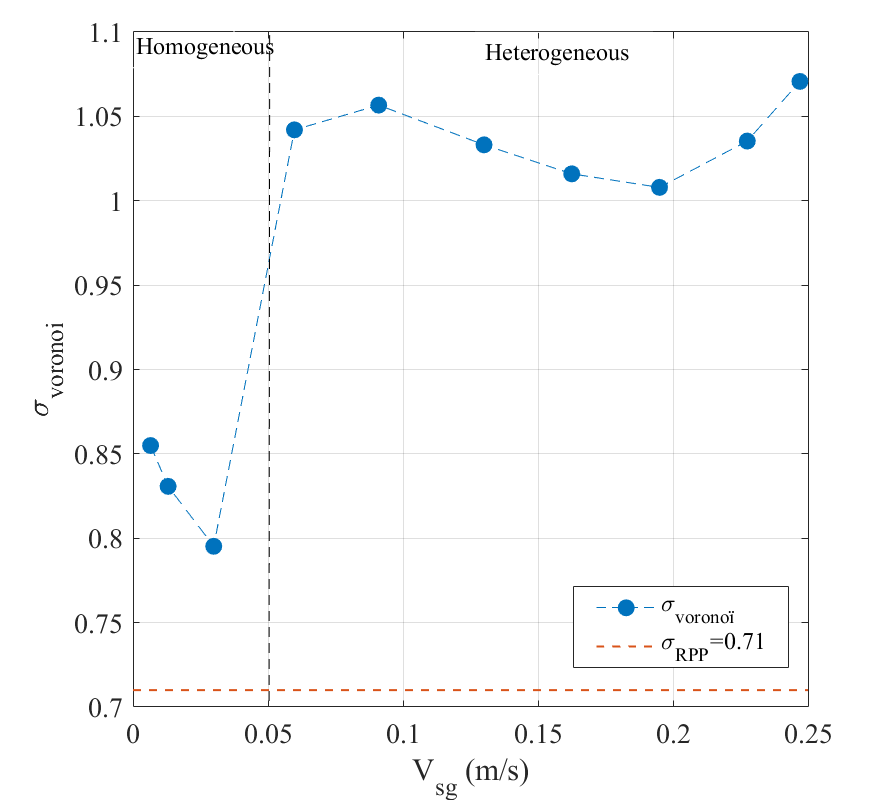}
         \caption{}
         \label{fig15b}
     \end{subfigure}
        \caption{(a) Centered pdfs of 1D Vorono\"i cells width $\Delta T_k/\langle\Delta T_k\rangle$ built from Doppler probe signals at various gas superficial velocities. The dash line represents the 1D Vorono\"i distribution for an RPP, i.e. a Random Poisson Process. The vertical dashed lines indicate the thresholds for the definitions of clusters and voids (as defined by \cite{monchaux2010preferential}). (b) Evolution of the standard deviation of 1D Vorono\"i distributions with the gas superficial velocity. The horizontal dash line indicates the standard deviation for an RPP while the vertical dashed line delineate the homogeneous-heterogeneous transition. Measurements performed in a $D=0.4$m column, on the column axis at $H/D=3.625$. }
        \label{fig15}
\end{figure}

Probability density functions (pdfs) of the Vorono\"i cell width $\Delta T_k$ normalized by the average $\langle \Delta T_k\rangle$ are presented in Figure \ref{fig15}a for various gas superficial velocities: all these data have been collected on the bubble column axis at $H/D=3.625$. Care was taken to ensure a correct convergence of these distributions. The latter comprises between 8000 and 13000 bubbles: these samples correspond to measuring durations from 95 to 950 seconds depending on flow conditions. 

Qualitatively, the width $\Delta T_k$ of the time interval containing the $k^{th}$ bubble is an indication of the local concentration. A short duration $ \Delta T_k$ means the presence of a close-by bubble while a large duration indicates that the $k^{th}$ bubble is somewhat isolated. We will come back later on the connection between normalised cell durations $\Delta T_k/\langle \Delta T_k\rangle$ and concentration. For the time being, let us focus on the allure of these pdfs. The dash line in Figure \ref{fig15}a represents the pdf of normalized cell durations $\Delta T_k/\langle \Delta T_k\rangle$ for a Random Poisson Process - RPP in short (\cite{ferenc2007size}) - that has no correlation at any scale. Clearly, and as noted by \cite{raimundo2019hydrodynamics}, measured distributions at large enough $V_{sg}$ differ from the RPP case. In particular, both very large cell durations (corresponding to dilute conditions) and very small cell durations (corresponding to dense conditions) are more probable than for a RPP.  

Following \cite{monchaux2010preferential}, the distance to a RPP is commonly appreciated by examining the standard deviation $\sigma_{voronoi}$ of the pdf of Vorono\"i cells widths. As shown in Figure \ref{fig15}b, such standard deviation drastically increases from a low value, comparable to that of a RPP, to a much higher value (close to unity) when the system shifts from the homogeneous to the heterogeneous regime. In the homogeneous regime, the measured standard deviation of Vorono\"i cells pdfs evolves between 0.8 and 0.85. This is slightly larger than the 0.71 limit for a RPP of point particles as determined by \cite{ferenc2007size} (according to \cite{uhlmann2020voronoi}, the standard deviation for a RPP with finite size particles is even lower). The origin of that small difference is unclear. That could be the mark of an inhomogeneous spatial repartition of bubbles in the homogeneous regime because of some gas maldistribution at injection (\cite{nedeltchev2020precise}): such a scenario is supported by the analysis of liquid velocity profiles (\cite{lefebvre2022new}). Alternately, that small difference could be due to the measuring method itself because the optical probe allows detecting the centers of gas chords and not the centers of bubbles, and because most bubbles are not spherical. Hence, the value of the standard deviation measured in the homogeneous regime could be interpreted at the reference RPP level as detected with the probe technique. The key points in Figure \ref{fig15}b are the very sharp increase in $\sigma_{voronoi}$ observed at the homogeneous-heterogeneous transition, and the large value, well above that of a RPP, that $\sigma_{voronoi}$ reaches at high $V_{sg}$. The shortcomings of 1D Vorono\"i analysis in complex flows (\cite{mora2018pitfalls,mora2019characterizing}) have been discussed elsewhere: one key result is that the neat difference observed with the standard deviation of a RPP unambiguously demonstrates that clustering does occur in the present flow conditions. Furthermore, for all heterogeneous conditions investigated (that is for $V_{sg}$ up to 24cm/s), the standard deviation $\sigma_{voronoi}$ remains nearly the same: that feature also indicates that clustering is a central characteristic of the heterogeneous regime. Finally, let us underline that, as for turbulent flows laden with inert particles (\cite{sumbekova2017preferential}), the main contribution to the standard deviation comes from cells at large $\Delta T_k/\langle \Delta T_k\rangle$ corresponding to low void fractions, compared with the contribution from cells with intermediate $\Delta T_k/\langle \Delta T_k\rangle$ (void fractions close to the mean value) or with low $\Delta T_k/\langle \Delta T_k\rangle$ (high void fractions).

To quantify the connection between cell width $\Delta T_k/\langle \Delta T_k\rangle$ and concentration, we consider two approaches. First, we follow what we did for turbulent flow laden with droplets (\cite{sumbekova2017preferential,mora2018pitfalls}), by connecting the ratio $\Delta T_k/\langle \Delta T_k\rangle$ with linear number densities, i.e. with the number of inclusions detected per unit length. The length corresponds to the measuring duration multiplied by the axial velocity $V_{axial}$ of inclusions. The local number density $\gamma_k$ (number of inclusions per meter) in the $k^{th}$ cell equals $1/[\Delta T_k V_{axial}] $ while $1/[\langle \Delta T_k\rangle V_{axial}] $ is the mean number density $\gamma$. Therefore, the normalized cell width $\Delta T_k/\langle \Delta T_k\rangle=\gamma / \gamma_k$ represents the inverse of the instantaneous (i.e. at the scale of the Vorono\"i cell) number density divided by the mean number density. When applied to bubble columns (\cite{raimundo2015analyse,mezui2018characterization,raimundo2019hydrodynamics}), we considered $V_{axial}$ as the mean bubble velocity, and $\gamma$ was assumed to be proportional to the mean dispersed phase concentration. Under these assumptions, the inverse of $\Delta T_k/\langle \Delta T_k\rangle$, i.e. $\gamma_k/\gamma$, provides the magnitude of the local gas concentration (local at the scale of the Vorono\"i cell) with respect to the mean gas fraction at the measuring location. In Figure \ref{fig15}a, the abscissa $\Delta T_k/\langle \Delta T_k\rangle$ varies from 0.07 to 10 so that $\gamma_k/\gamma$ covers more than two decades as it evolves between 0.1 and about 14.

However, a second approach is required because, for the heterogeneous conditions considered here, $\gamma_k/\gamma$ does not coincide with the ratio $\varepsilon_k / \varepsilon$ of the void fraction $\varepsilon_k$ relative to the $k^{th}$ cell to the mean gas hold-up $\varepsilon$ at the measuring location. Indeed, in the turbulent laden flows we have previously analysed, all inclusions traveled with almost the same axial velocity. This is no longer the case for bubbles in the heterogeneous regime as their velocities experience strong variations (see figure 2 in \cite{Mezui2022}), leading to a standard deviation as large as 60\% of the mean. Hence, the selection of a mean bubble velocity to transform time into space induces very large distorsions on the concentration estimate by way of $\gamma_k$. To correct for these distortions and to evaluate reliable local void fractions, it is appropriate to rely on gas residence times as the latter naturally account for the actual velocity of each bubble. The void fraction relative to the $k^{th}$ Vorono\"i cell equals the sum of gas residence times included in that cell divided by the cell duration $\Delta T_k$. As shown in appendix \ref{appB}, the ratio $\Delta T_k/\langle \Delta T_k\rangle$ is indeed related with $\varepsilon/\varepsilon_k$, but it does not coincides with $\varepsilon/\varepsilon_k$ as the prefactor between these two quantities varies with the gas residence time (see equation \ref{eqapp} in appendix \ref{appB}). In the following, we will use the ratio $\gamma_k/\gamma$ as a crude, qualitative characterisation of meso-scale structures in terms of concentration, while exact measurements of the gas fraction $\varepsilon_k / \varepsilon$ will be considered in Section \ref{sec5} for discussing modelling issues.

Going back to Figure \ref{fig15}a, and whatever the flow conditions, the measured pdfs of Vorono\"i cells cross the RPP at two fixed abscissa represented by vertical dashed lines. A third intersection sometimes occurs in the very dense limit (at $\Delta T_k/\langle\Delta T_k\rangle$ about 0.1, that is for $\gamma_k/\gamma$ about 10), but it will not be considered here because its occurrence is far too sensitive to the sample size. As for turbulent laden flows (\cite{monchaux2010preferential}), we define three populations out of the two stable thresholds. A Vorono\"i cell (and the bubble it contains) belongs to a `dense' region when $\Delta T_k/\langle\Delta T_k\rangle$ is below 0.51, or equivalently when $\gamma_k/\gamma$ is higher than 1.96. A Vorono\"i  cell (and the bubble it contains) belongs to an `empty' or `void' region when $\Delta T_k/\langle\Delta T_k\rangle$ is above 2.89, or equivalently when $\gamma_k/\gamma$ is lower than 0.34. In between, the cell (and its bubble) pertains to an `intermediate' region. Owing to Figure \ref{fig15}a, the probability for bubbles to belong to `empty' or to `dense' regions is larger than in RPP. This is confirmed by the data presented in Table \ref{tab3}. In average, 38\% of the bubbles belong to dense regions, while 5\% are within empty regions and 57\% are in intermediate regions: these figures remain stable within about 5\% over the whole heterogeneous regime that is for $V_{sg}$ from 6cm/s to 25cm/s. As expected, the figures relative to dense and to empty regions are significantly larger than those for a RPP. Table \ref{tab3} also confirms that, in the homogeneous regime, the repartition of bubbles in number between dense, empty and intermediate regions is very close to the repartition in number for an RPP. 

The difference between homogeneous and heterogeneous conditions is also manifest in terms of void fractions. In the heterogeneous regime, the contributions to the local void fraction are typically 17\% for the dense regions, 10\% for the empty regions and 70\% for intermediate regions (Table \ref{tab3}). These values correspond to average values for $V_{sg} \geq 9 $ cm/s: they change by less than 1\% when considering data over the interval $V_{sg} \geq 6 $ cm/s. Note that there is a slight decrease of the contribution of dense regions to the local void fraction as $V_{sg}$ increases, which is compensated by a slight increase with $V_{sg}$ of the contributions of empty and intermediate regions. 

\begin{table}
\centering
\resizebox{\textwidth}{!}{%
\begin{tabular}{|l|c|c|c|}
\hline
 & Dense regions & Intermediate regions & Empty regions \\ \hline
\cellcolor[HTML]{FFFFFF}\begin{tabular}[c]{@{}l@{}}Repartition of   bubbles in number in measured pdfs\\  in the heterogeneous regime\\  (average   values for $V_{sg}\geq$  9cm/s)\end{tabular} & 38\% & 57\% & 5\% \\ \hline
\begin{tabular}[c]{@{}l@{}}Repartition of the actual void fraction $\varepsilon_k/\varepsilon$ \\ in  the heterogeneous regime (average for $V_{sg} \geq$ 9cm/s)\end{tabular} & 17\% & 73\% & 10\% \\ \hline
\cellcolor[HTML]{FFFFFF}Repartition of bubbles in number in RPP & 30\% & 68\% & 2\% \\ \hline
\cellcolor[HTML]{FFFFFF}\begin{tabular}[c]{@{}l@{}}Repartition of   the actual void fraction $\varepsilon_k/\varepsilon$ \\ in the homogeneous regime (average values for $V_{sg}\leq $ 3cm/s) \end{tabular} & \cellcolor[HTML]{FFFFFF}29\% & \cellcolor[HTML]{FFFFFF}67\% & 4\% \\ \hline
\cellcolor[HTML]{FFFFFF}\begin{tabular}[c]{@{}l@{}}Repartition of bubbles in number   \\ in measured pdfs in the homogeneous regime \\ (values for $V_{sg}$   = 1.3cm/s)\end{tabular} & 33.60\% & 63.20\% & 3.20\% \\ \hline
\end{tabular}%
}
\caption{Typical distributions of the dispersed phase between void regions, intermediate regions and dense regions in the heterogeneous and homogeneous regimes and comparison with a RPP. From measurements on the axis of a $D=0.4$m bubble column at $H/D=3.625$. }
\label{tab3}
\end{table}

\subsection{Characterisation of meso-scale structures \label{sec51}}

\begin{figure}
     \centering
     \begin{subfigure}[b]{0.5\textwidth}
         \centering
         \includegraphics[width=\textwidth]{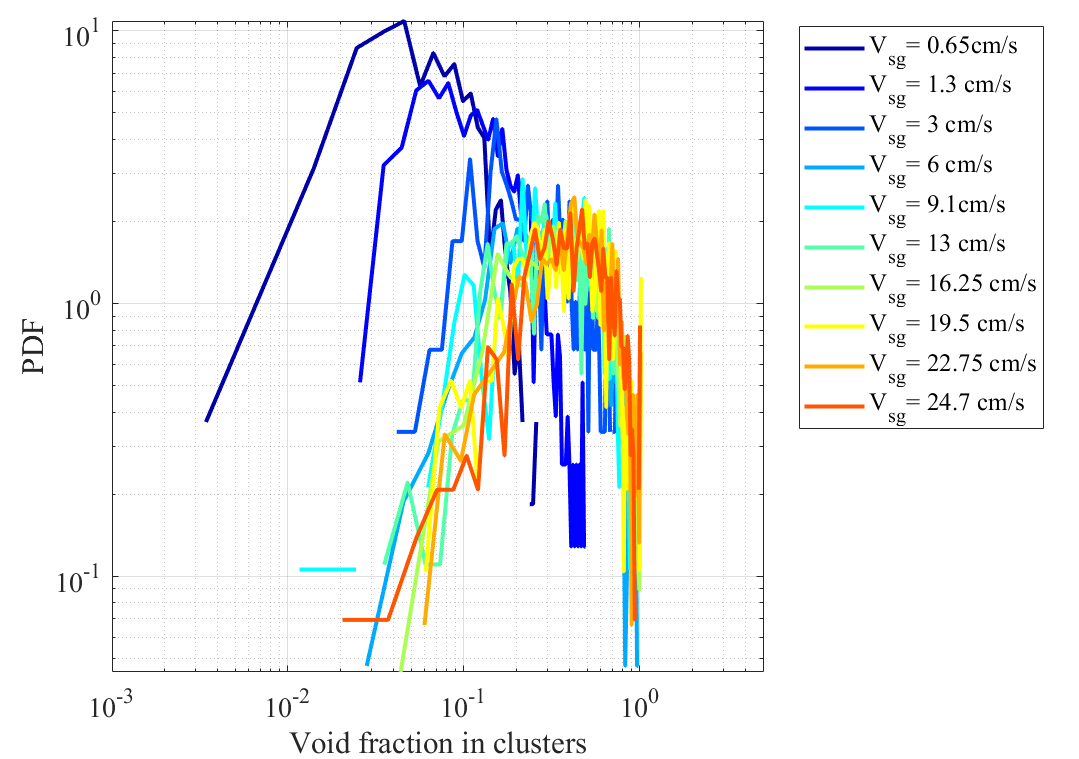}
         \caption{}
         \label{fig16a}
     \end{subfigure}
     \hfill
     \begin{subfigure}[b]{0.375\textwidth}
         \centering
         \includegraphics[width=\textwidth]{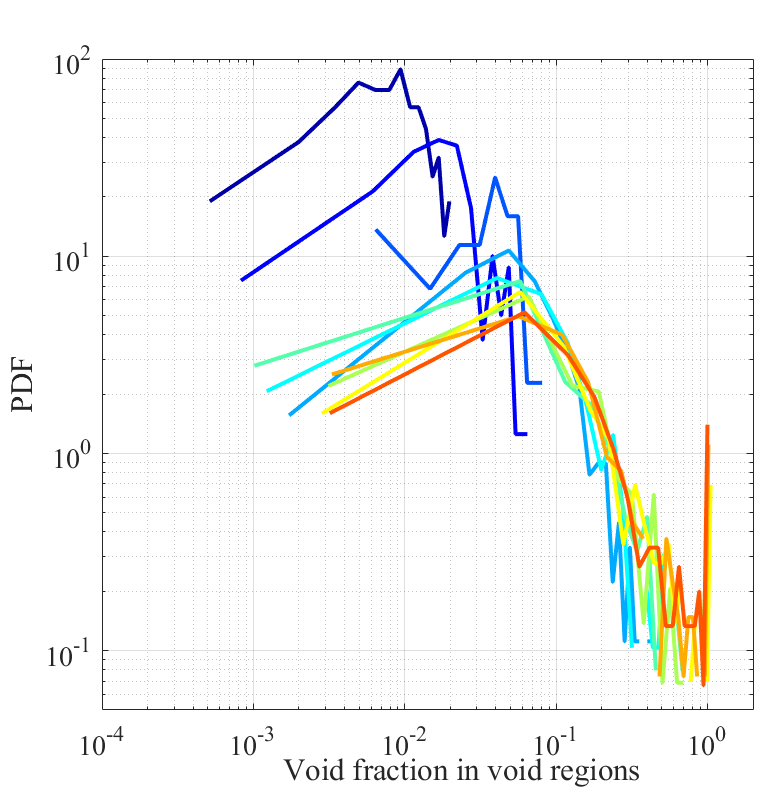}
         \caption{}
         \label{fig16b}
     \end{subfigure}
        \caption{Pdfs of void fraction (in absolute value) in clusters (a) and in void regions (b) for different superficial velocities. For these statistics, we considered clusters comprising at least two bubbles. Measurements performed in a $D=0.4$m column, on the column axis at $H/D=3.625$.}
        \label{fig16}
\end{figure}

\begin{figure}
     \centering
     \begin{subfigure}[b]{0.45\textwidth}
         \centering
         \includegraphics[width=\textwidth]{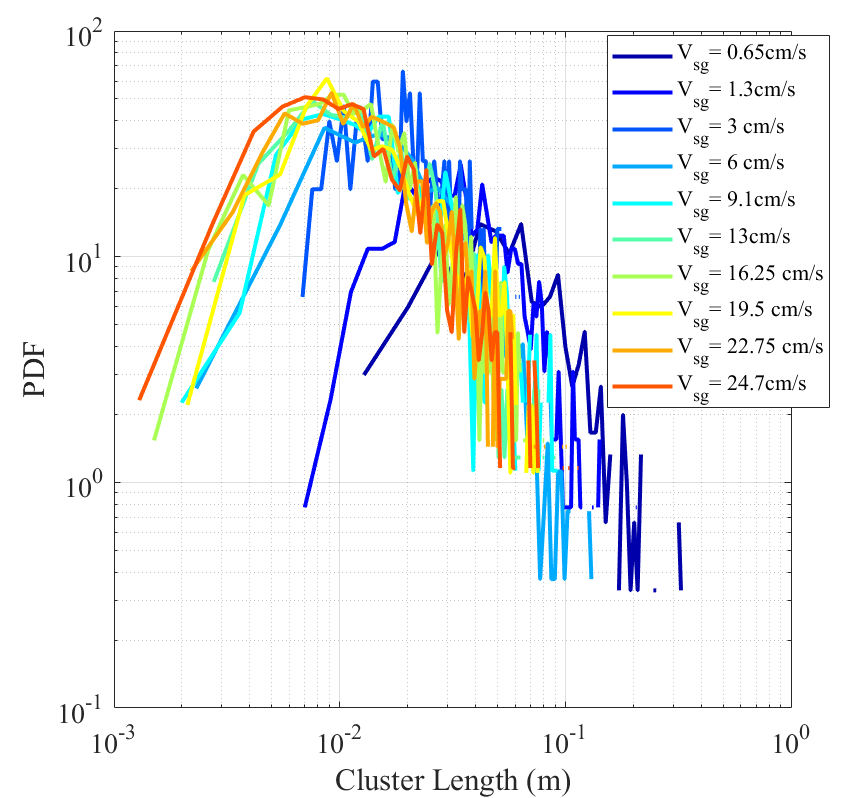}
         \caption{}
         \label{fig17a}
     \end{subfigure}
     \hfill
     \begin{subfigure}[b]{0.47\textwidth}
         \centering
         \includegraphics[width=\textwidth]{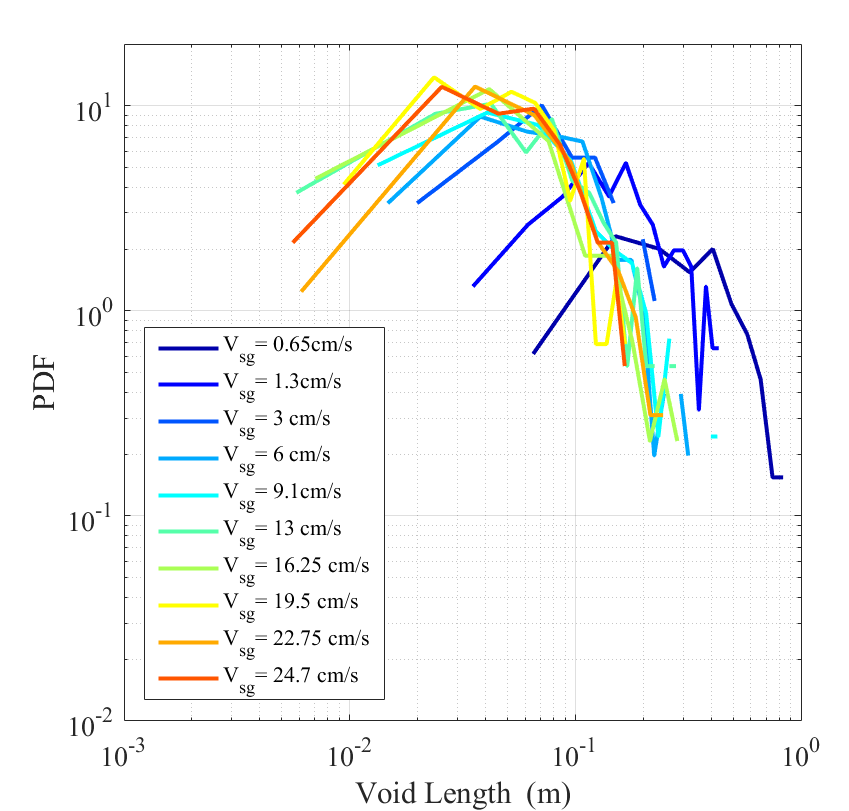}
         \caption{}
         \label{fig17b}
     \end{subfigure}
        \caption{(a) Pdfs of lengths of clusters (a) and of void regions (b) for different superficial velocities. For these statistics, we considered clusters that comprise at least two bubbles. Measurements performed in a $D=0.4$m column, on the column axis at $H/D=3.625$.}
        \label{fig17}
\end{figure}

Once all bubbles have been distributed within the three populations, meso-scale structures are then formed using the following procedure. Bubbles belonging to a `dense' region and successive in time are assembled to form a `cluster'. Similarly, successive bubbles belonging to an `empty' region are assembled to form a `void'. The same process was used for intermediate regions. The characteristics of the resulting meso-scale structures in terms of size and concentration are then extracted:

\begin{itemize}

\item The void fraction (in absolute value) in a given meso-scale structure is evaluated as the sum of gas residence times for all bubbles pertaining to that structure divided by the duration of that structure, the later being the sum of all involved $\Delta T_k$. The distributions of void fraction in clusters and in voids are exemplified in Figure \ref{fig16} for various $V_{sg}$.

\item The size of a given meso-scale structure is estimated as the duration of the structure multiplied by the average bubble velocity, the latter being evaluated for the bubbles belonging to the structure considered: these conditional velocities are analyzed in the next section. Length distributions for clusters and for voids are provided Figure \ref{fig17} for various $V_{sg}$.

\end{itemize}

We considered two options for clusters: either the minimum number of bubbles in a cluster is set to 1 so that all Vorono\"i cells with a $\Delta T_k/\langle\Delta T_k\rangle$ below the threshold are considered as clusters, or the minimum number of bubbles is set to 2 so that clusters involving a single bubble are excluded. That second option has been suggested to help distinguishing between `coherent' and `random' clusters in turbulent laden flows (\cite{mora2019characterizing}). Here, and for all the flow conditions pertaining to the heterogeneous regime, it happens that 37\% to 40\% of clusters involve a single inclusion.
 
 It should also be underlined that zones below or beyond the above-defined thresholds also exist for a RPP. Hence, one can still identify and statistically characterize `dilute' and `dense' regions in homogeneous conditions even though the corresponding Vorono\"i distributions are very close to and/or almost collapse with a RPP. Using the same data processing routine to analyze homogeneous and heterogeneous conditions, the characteristics of clusters and of empty regions are presented over the whole range of $V_{sg}$ from homogeneous to heterogeneous regimes, bearing in mind that different physical origins are associated with meso-scale structures for these two regimes. In particular, the data in the homogeneous regime are not expected to bear any particular significance as they could be of random origin, or they could be related to some correlation induced by `defects' in the system (due for example to gas injection, see \cite{lefebvre2022new}). 

\begin{figure}
     \centering
     \begin{subfigure}[b]{0.48\textwidth}
         \centering
         \includegraphics[width=\textwidth]{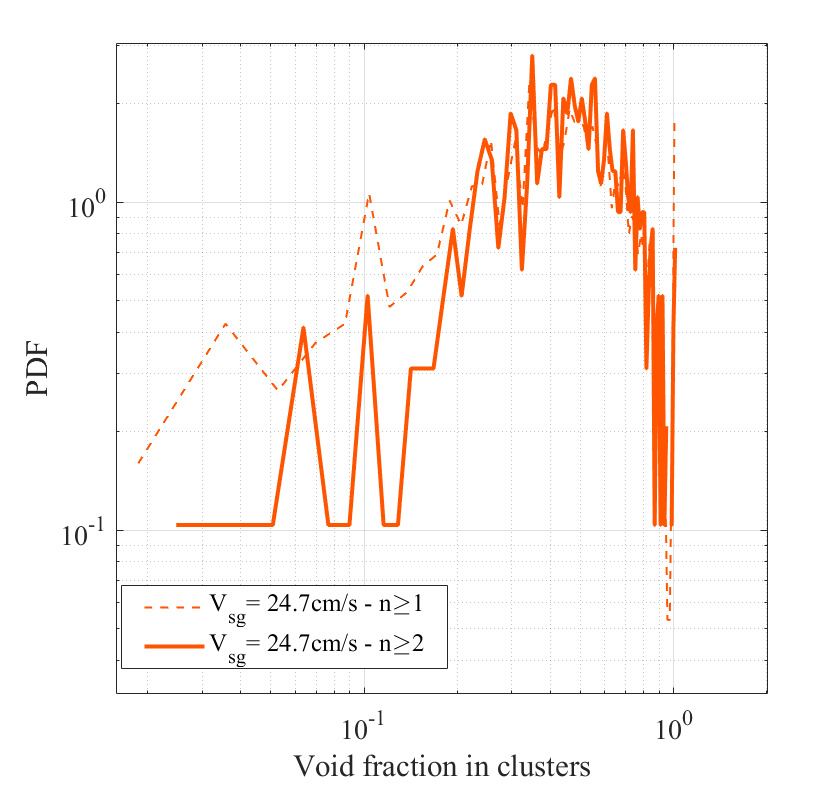}
         \caption{}
         \label{fig18a}
     \end{subfigure}
     \hfill
     \begin{subfigure}[b]{0.47\textwidth}
         \centering
         \includegraphics[width=\textwidth]{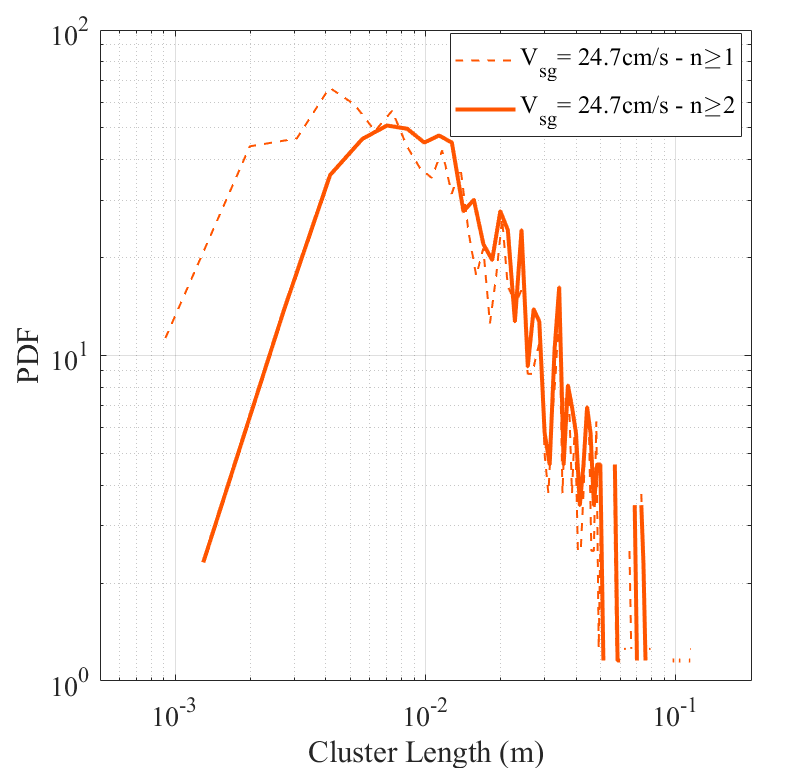}
         \caption{}
         \label{fig18b}
     \end{subfigure}
        \caption{Comparisons of the pdfs of void fraction in clusters (a) and of clusters lengths (b) when the minimum number of bubbles is set to 1 or to 2. Measurements performed in a $D=0.4$m column, on the column axis at $H/D=3.625$ and $V_{sg}=24.7$cm/s. }
        \label{fig18}
\end{figure}

Figures \ref{fig16} and \ref{fig17} clearly demonstrate that, for void regions as well as for clusters, the distributions in the heterogeneous regime markedly differ from the distributions observed in the homogeneous regime. Moreover, in the heterogeneous regime, the distributions tend to collapse indicating that clusters and void regions reach an asymptotic state when the gas superficial velocity becomes large enough. As shown in Figure \ref{fig18}, that limiting state is almost the same when considering clusters with a minimum of 1 bubble or with a minimum of 2 bubbles.

\begin{figure}
\centering
\includegraphics[width=\textwidth]{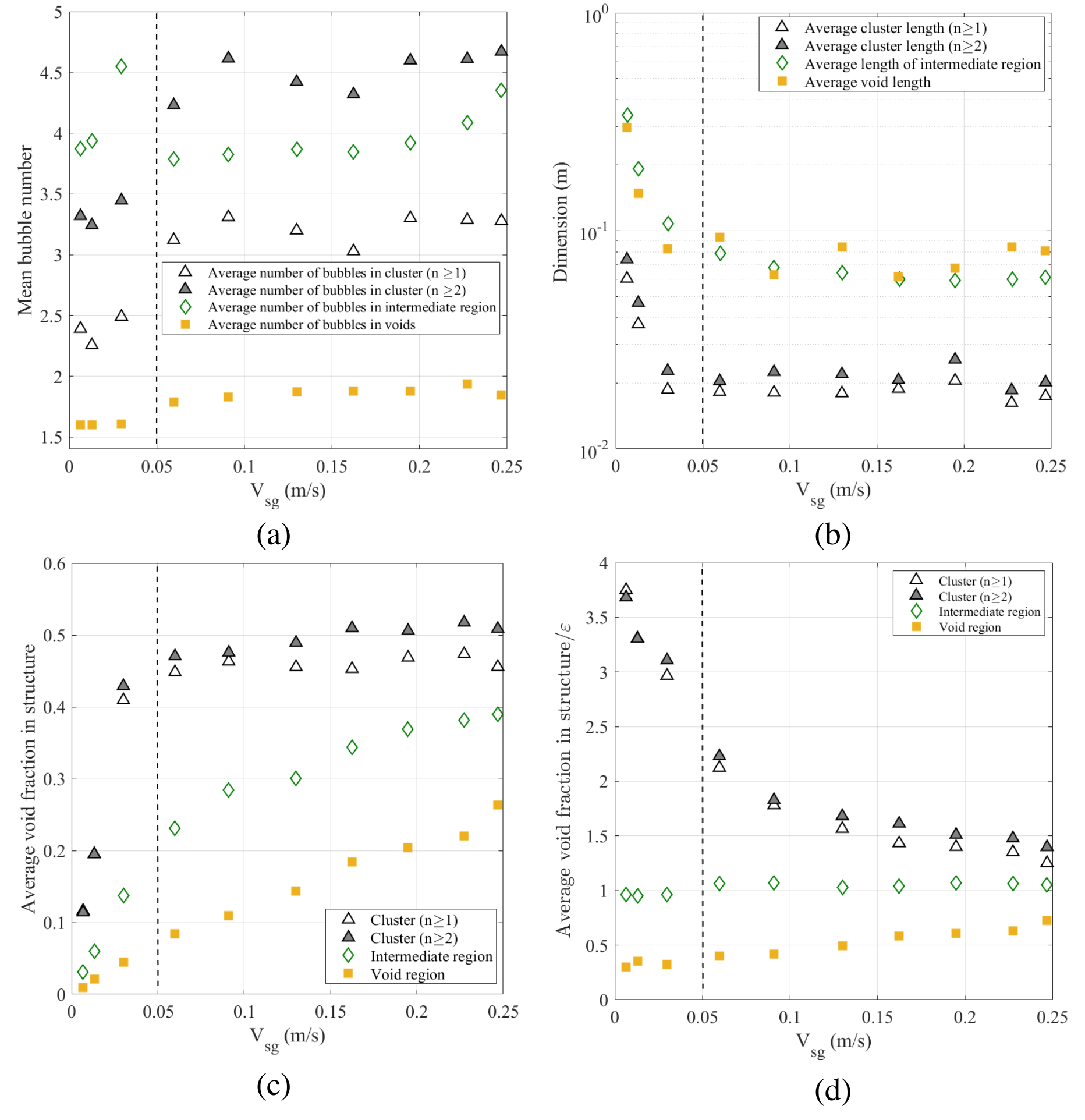}
\caption{Mean characteristics of clusters, of void regions and of intermediate regions versus the gas superficial velocity: (a) Average number of bubbles in meso-scale structures, (b) average size, (c) average absolute gas concentration in meso-scale structures, (d) average concentration scaled by the void fraction on the column axis. Measurements performed in a $D=0.4$m column, on the column axis, at $H/D=3.625$. Vertical dashed lines delineate the homogeneous to heterogeneous transition. } \label{fig19}
\end{figure}

The average characteristics of clusters and of void and intermediate regions are given Figure \ref{fig19} as a function of the gas superficial velocity. 

\begin{itemize}

\item The average number of bubbles is about 1.8 in void regions and about 4 in intermediate regions. In clusters, it is about 4.5 when $n
\geq2$, and it drops to 3.2 when accounting for clusters consisting of a single bubble. The decrease from 4.5 to 3.2 is consistent with the fact that, as seen above, 2/3 of the clusters comprise more than one bubble. These average numbers of bubbles are quite low: they indicate that the clusters are not organized as compact assemblies of bubbles, but are more like thin sheets. The fact that the probability to find a cluster comprising $N$ bubbles decays like $N^{-1.17}$, i.e. that it strongly drops with $N$, also supports the proposed picture. In particular, 1D clusters comprising more than 10 bubbles are very rare: they represent only 3.7\% of the clusters (with $n\geq1$) present in the heterogeneous regime.

\item The size of void regions and of intermediate regions varies from 6-7cm to 20cm while the size of clusters ranges from a few millimeters up to $6-7$cm. In the heterogeneous regime, the mean size of clusters $\langle L_{cluster} \rangle$, that of void regions $\langle L_{void} \rangle$ and that of intermediate regions $\langle L_{int} \rangle$ all remain fairly stable. The mean cluster length asymptotes at $21\pm3mm$: it is marginally affected if one considers a minimum of one bubble instead of two to form clusters. The asymptotic mean length of void regions is significantly larger as $\langle L_{void} \rangle\rangle \sim 74$mm$\pm10$mm, and similarly, for intermediate regions, $\langle L_{int} \rangle$ is about $62$mm$\pm4$mm. 
 
\item The average concentration (in absolute value) in voids steadily increases with the gas superficial velocity. A similar behavior holds for intermediate regions. In clusters, the average concentration sharply increases at the homogeneous-heterogeneous transition, and for $V_{sg}$ above $\sim0.15$m/s, it tends to stabilize at a large void fraction, say about 50\%. Interestingly, when scaled by the local void fraction $\varepsilon$ (here $\varepsilon$ equals the void fraction on the axis $\varepsilon_{axis}$), the mean concentrations in voids and in intermediate regions increase with the mean gas hold-up, while the concentration in clusters slightly decreases: additional data are needed to confirm if the asymptotic trend corresponds to a decrease or to a plateau. The same question holds concerning the asymptotic behavior of the difference in concentration between dense and dilute regions.
\end{itemize}

\section{Absolute and relative bubble velocities conditioned by the local concentration \label{sec4}}

\begin{figure}
     \centering
     \begin{subfigure}[b]{0.47\textwidth}
         \centering
         \includegraphics[width=\textwidth]{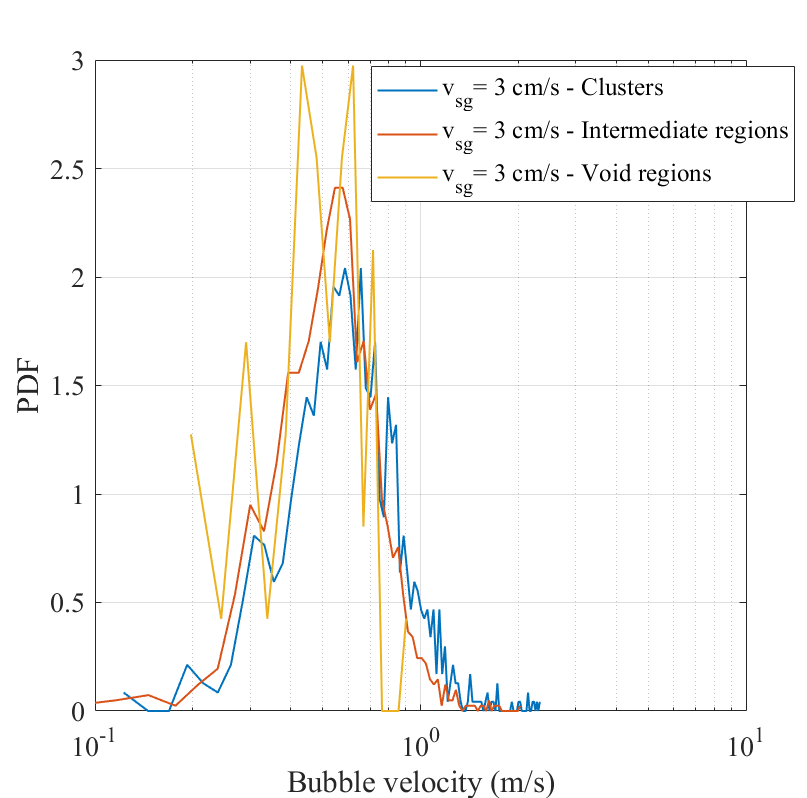}
         \caption{}
         \label{fig21a}
     \end{subfigure}
     \hfill
     \begin{subfigure}[b]{0.47\textwidth}
         \centering
         \includegraphics[width=\textwidth]{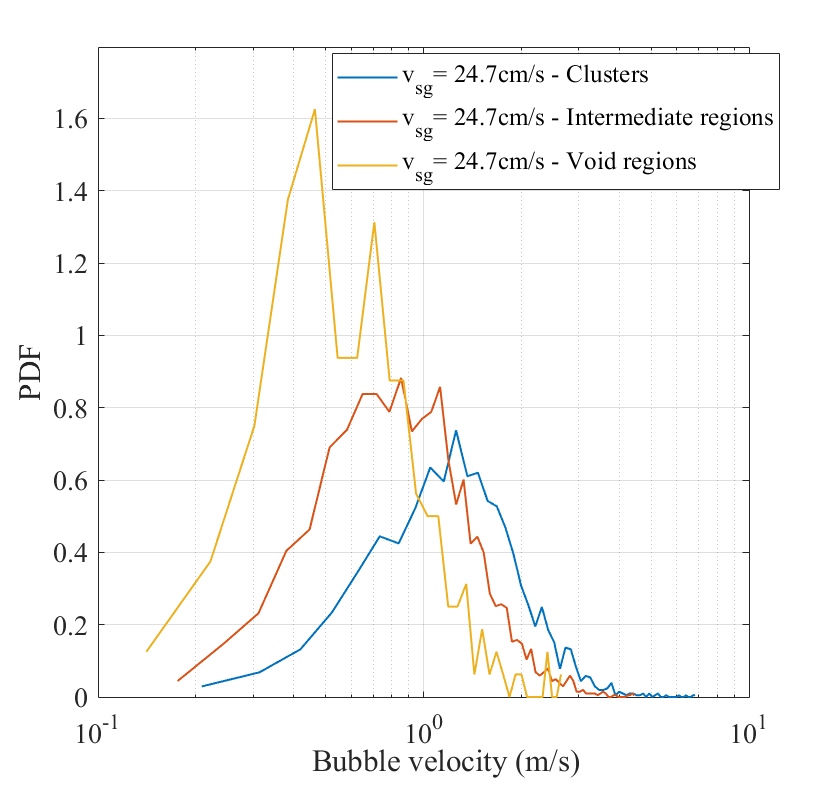}
         \caption{}
         \label{fig21b}
     \end{subfigure}
        \caption{Bubble velocity pdfs conditioned by the meso-scale structure they belong to, i.e. clusters, intermediate regions or void regions for $V_{sg}=3$cm/s (a) and for $V_{sg}=24.7$cm/s (b). Measurements performed in a $D=0.4$m column, on the column axis at $H/D=3.625$ with a downward directed Doppler probe.}
        \label{fig21}
\end{figure}

Paralleling what we did for turbulent laden flows (\cite{sumbekova2016clustering}), bubbles are classified into three populations namely clusters, void regions and intermediate regions. Bubble velocity pdfs are built for each of these populations using direct velocity measurements (no interpolation) performed with a downward oriented Doppler probe (\cite{lefebvre2022new}). Examples of such conditional pdfs are provided Figure \ref{fig21}. For both regimes, the minimum velocities are about the same for the three populations, while the most probable velocity as well as the maximum velocity drift to larger values when successively considering void regions, intermediate regions and clusters. This drift is weak in the homogeneous regime: the velocity at the peak increases from about 0.4m/s in void regions to 0.7m/s in clusters, so that the difference is of the order of the bubble terminal velocity. The drift is significantly larger in the heterogeneous regime as the most probable velocity goes from $\sim 0.5$m/s in void regions up to 1.3m/s in clusters: in that case, the difference amounts to 3.5 times the bubble terminal velocity. Hence, the conditional bubble velocities gathered with the Doppler optical probe confirm our physical expectation that, in average, high void fraction regions are moving up much faster than low void fraction regions.

\begin{figure}
\centering
\includegraphics[width=0.85\textwidth]{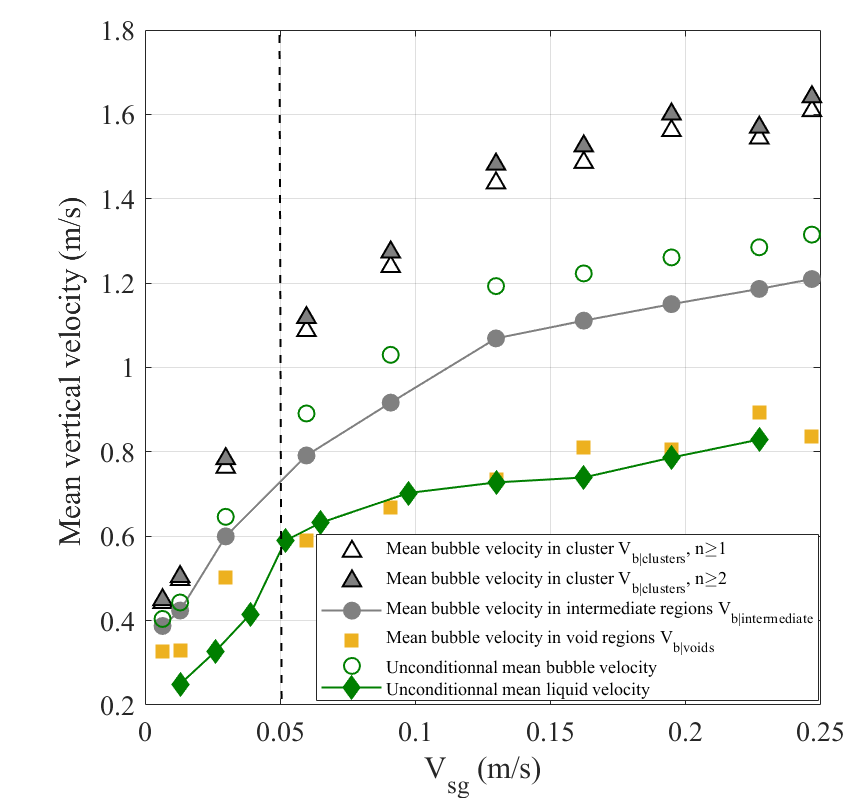}
\caption{Average absolute bubble velocity for bubbles pertaining to clusters $V_{b \vert clusters}$, to intermediate regions $V_{b \vert int}$ and to void regions $V_{b \vert voids}$ versus the gas superficial velocity. Measurements performed in a $D=0.4$m column, on the column axis at $H/D=3.625$ with a downward directed Doppler probe. The unconditional mean liquid $V_L$ and gas $V_G$ velocities from Figure \ref{fig:fig1} are also shown for sake of comparison.} \label{fig22}
\end{figure}

To quantify this effect, and for each meso-scale structure, we evaluated the mean bubble velocity $V_{b \vert s}$ for bubbles pertaining to the selected meso-scale structure. These velocities, that represent absolute velocities in the laboratory frame, are shown in Figure \ref{fig22} as a function of the gas superficial velocity. It could be observed that the average conditional velocities relative to void regions $V_{b \vert voids}$, to intermediate regions $V_{b \vert int}$ and to clusters $V_{b \vert cluster}$, all monotonously increase with $V_{sg}$. Beside, the velocity differences between any two out of these three populations remain limited, in the order of $U_T$, in the homogeneous regime. Beyond the homogeneous-heterogeneous transition, the velocity differences neatly increase with $V_{sg}$: bubbles embedded in dense regions are moving up faster than bubbles in intermediate regions, which are themselves moving up faster than bubbles in dilute regions. This observation provides an indisputable evidence of the central role of meso-scale structures on the actual dynamics of bubbles in the heterogeneous regime.

In Figure \ref{fig22}, we have reported the unconditional mean vertical bubble velocity $V_G$ shown in Figure \ref{fig:fig1} (green dots). In the heterogeneous regime, $V_G$ happens to be comprised between $V_{b\vert int}$ and $V_{b \vert cluster}$. It is tempting to try to recover the unconditional bubble velocity $V_G$ from conditional measurements. Considering that a fraction $N_{cluster}$ of bubbles pertains to clusters, that a fraction $N_{int}$ belongs to intermediate regions and a fraction $N_{void}$ to void regions (with $N_{cluster} + N_{int} + N_{void} = 1$), one expects that:

\begin{equation}\label{eqN1}
V_G = N_{cluster} V_{b \vert cluster} + N_{int} V_{b \vert int} + N_{void} V_{b \vert void}.					
\end{equation}

For the heterogeneous regime, the conditional mean bubble velocities $V_{b \vert cluster}$, $V_{b \vert int}$  and $V_{b \vert void}$ are provided in Table  \ref{tab1NN} for all the gas superficial velocities considered in the experiments. As the repartition of bubbles between the three populations is already known (see Table \ref{tab3}), the mean bubble velocity deduced from equation \ref{eqN1} can be evaluated. As shown in Table \ref{tab1NN}, there is an excellent agreement between the bubble velocity predicted using equation \ref{eqN1} and direct, unconditional measurements of the bubble velocity. Moreover, the contribution of bubbles inside clusters to their vertical transport velocity amounts to 46\%, the contribution of intermediate regions is 51\% and the remaining 3\% arise from void regions. These figures remain the same within $\pm0.5$\% for all $V_{sg}$ considered in the heterogeneous regime. 

These results provide more evidence that meso-scale structures have a key role in the dynamics of bubbles and notably on their absolute velocity in the heterogeneous regime. In some way, they confirm the intuition of Noel De Nevers concerning the role of internal structures, as this author argued in 1968: \textit{``In unbaffled systems these (bubble driven) circulations are unstable and chaotically change in size, shape, and orientation. These chaotic circulations provide the principal mode of vertical bubble transport in bubble columns over a wide range of operating conditions''} (\cite{de1968bubble}). 

\vspace{0.25cm}
\begin{adjustwidth}{-2.5 cm}{-2.5 cm}\centering\begin{threeparttable}[!htb]
\scriptsize
\begin{tabular}{|l|r|r|r|r|r|r|r|}\toprule
&\multicolumn{3}{|c|}{Conditional mean bubble velocities (m/s)} &\multicolumn{2}{|c|}{Unconditional mean bubble velocities (m/s)} & \\\cmidrule{2-6}
$V_{sg}$(m/s) &\makecell{Bubbles in \\ clusters} &\makecell{Bubbles in \\ intermediate regions} &\makecell{Bubbles in \\ void regions} & \makecell{Deduced from \\ equation \ref{eqN1}} &Measured &difference \% \\\midrule
\hline
0.091 &1.239 &0.917 &0.667 &1.027 &1.03 &-0.4 \\
\hline
0.13 &1.437 &1.069 &0.734 &1.192 &1.193 &0 \\
\hline
0.1625 &1.486 &1.111 &0.812 &1.238 &1.223 &1.3 \\
\hline
0.195 &1.561 &1.15 &0.805 &1.289 &1.2613 &2.2 \\
\hline
0.2275 &1.544 &1.186 &0.894 &1.307 &1.285 &1.8 \\
\hline
0.247 &1.609 &1.21 &0.836 &1.343 &1.3155 &2 \\
\hline
\hline
\makecell{Repartition of \\ bubbles in  \\ number (as measured)}  &38\% &57\% &5\% & & &    \\
\bottomrule
\end{tabular}
\caption{Estimations of the unconditional mean bubble velocity $V_G$ deduced from mean bubble velocities conditioned by meso-scale structures using equation \ref{eqN1} in the heterogeneous regime. }\label{tab1NN}
\end{threeparttable}\end{adjustwidth}
\vspace{0.25cm}

The same analysis was also done for the homogeneous regime. For each of the four gas superficial velocities $V_{sg}$ considered in that regime, Table \ref{tab2NN} provides the conditional bubble velocities, the resulting unconditional bubble velocity predicted using equation \ref{eqN1} and the unconditional bubble velocity that was directly measured. The agreement is very good, except for a 15\% difference for one condition. Note that very similar figures would be obtained if one considers the repartition for a RPP instead of the repartition of bubbles that was measured in the homogeneous regime (Table \ref{tab3}). Therefore, equation equation \ref{eqN1} allows to recover the unconditional bubble velocity from conditional data in the homogeneous regime. However, the contributions of each population to the vertical transport velocity of bubbles are different from those found in the heterogeneous regime: they amount to about 39\% for clusters, 59\% for intermediate regions and 2\% for void regions. 

\vspace{0.25cm}
\begin{adjustwidth}{-2.5 cm}{-2.5 cm}\centering\begin{threeparttable}[!htb]
\scriptsize
\begin{tabular}{|l|r|r|r|r|r|r|r|}\toprule
&\multicolumn{3}{|c|}{Conditional mean bubble velocities (m/s)} &\multicolumn{2}{|c|}{Unconditional mean bubble velocities (m/s)} & \\\cmidrule{2-6}
$V_{sg}$(m/s) &\makecell{Bubbles in \\ clusters} &\makecell{Bubbles in \\ intermediate regions} &\makecell{Bubbles in \\ void regions} & \makecell{Deduced from \\ equation \ref{eqN1}} &Measured &difference \% \\\midrule
\hline
0.0065 &0.442 &0.387 &0.327 &0.404 &0.404 &0.04 \\
\hline
0.013 &0.497 &0.424 &0.329 &0.446 &0.442 &0.6 \\
\hline
0.0299 &0.763 &0.600 &0.503 &0.652 &0.563 &14.5 \\
\hline
0.0598 &1.087 &0.791 &0.590 &0.884 &0.891 &-0.8 \\
\hline
\hline
\makecell{Repartition of \\ bubbles in  \\ number (as measured)}  &33.6\% &63.2\% &3.2\% & & &    \\
\bottomrule
\end{tabular}
\caption{Estimates of the unconditional mean bubble velocity $V_G$ deduced from mean bubble velocities conditioned by meso-scale structures using equation \ref{eqN1} in the homogeneous regime.}\label{tab2NN}
\end{threeparttable}\end{adjustwidth}
\vspace{0.25cm}

\begin{figure}
\centering
\includegraphics[width=\textwidth]{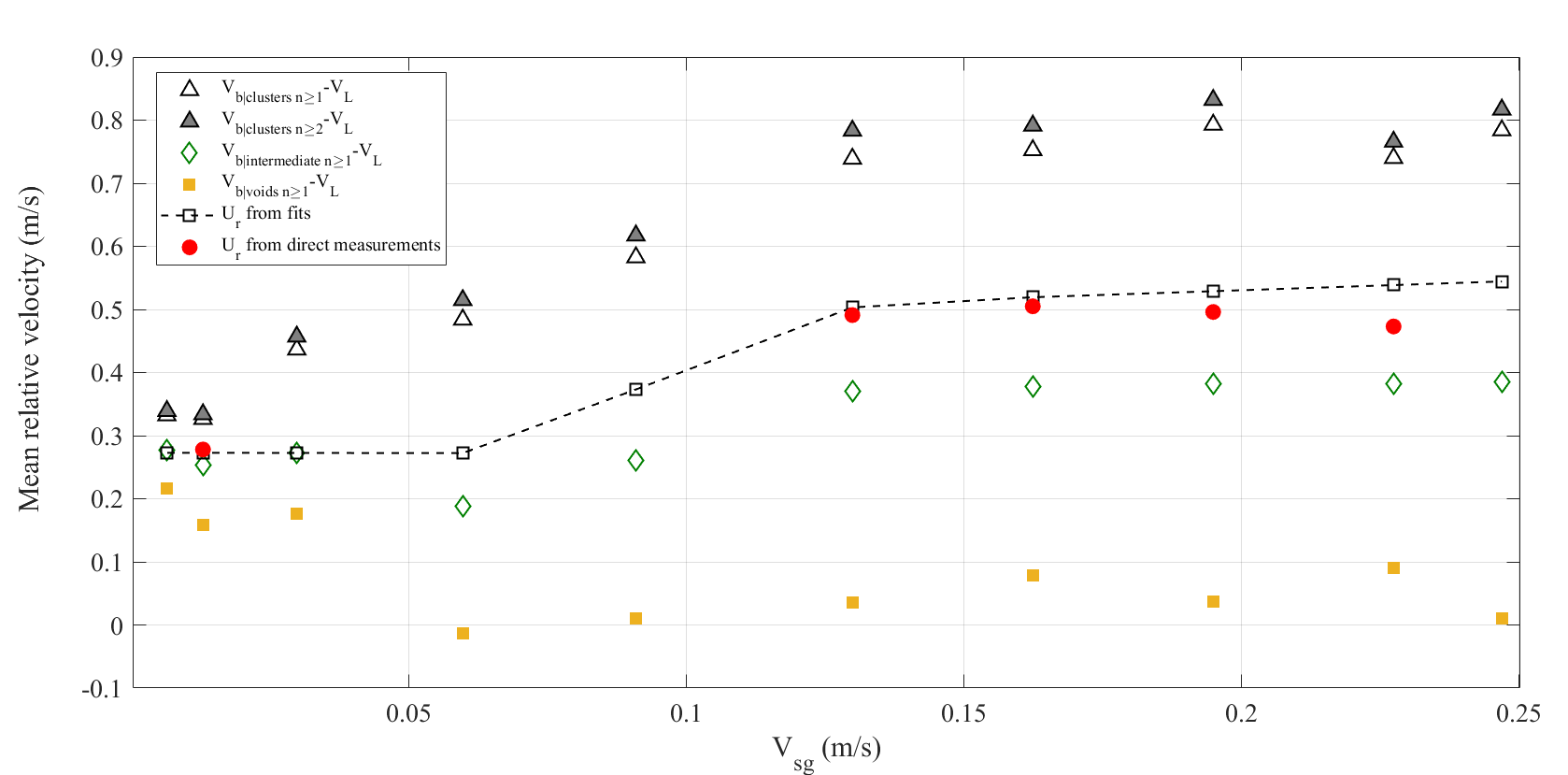}
\caption{Mean relative velocity between bubbles pertaining to a meso-scale structure, namely clusters, intermediate regions and void regions, and the liquid phase: evolution with $V_{sg}$. The unconditional relative velocity $U_R$ is also shown for comparison (red dots correspond to direct measurements while the dash line corresponds to interpolated data shown in Figure \ref{fig:fig1}). Measurements performed in a $D=0.4$m column, on the column axis at $H/D=3.625$. } \label{fig23}
\end{figure}

To appreciate the role of meso-scale structures on the relative motion, we plot in Figure \ref{fig23} the mean bubble relative velocities with respect to the liquid phase for each meso-scale structure. As shown in Figures \ref{fig22} and \ref{fig23} , the mean, unconditional liquid velocity is very close to the mean bubble velocity on void regions so that $V_{b \vert void} - V_L$ remains close to zero in the heterogeneous regime. This is not too surprising because void regions contains few bubbles, and also because of the bubble response time compared with its transit time through the column (see the discussion in Section \ref{sec6}). The two velocity differences $V_{b \vert clusters} - V_L$  and $V_{b \vert int} - V_L$ increase with $V_{sg}$ in a way similar to the unconditional relative velocity $U_R = V_G - V_L$. In particular, the differences in velocities remain moderate in the homogeneous regime, and they steeply increase at the transition. Both differences $V_{b \vert clusters} - V_L$  and $V_{b \vert int} - V_L$ tend to become more or less constant at large $V_{sg}$ (roughly above $V_{sg} \approx 13-15$cm/s). In intermediate regions, the average bubble velocity exceeds that of the liquid by $0.3-0.4$m/s. In clusters, the difference reaches about $0.7-0.8$m/s that is 3 to 3.5 times the bubble terminal velocity. 

A decomposition similar to equation \ref{eqN1} can be applied to the mean bubble relative velocity, namely:

\begin{equation}\label{eq11}
\begin{split}
U_R = V_G - V_L= N_{clusters} (V_{b \vert clusters}-V_L)+N_{int} (V_{b \vert int}-V_L)+\\N_{voids} (V_{b \vert voids}-V_L).
\end{split}
\end{equation} 

Even though equation \ref{eq11} can be directly deduced from equation \ref{eqN1}, it is important to analyse the contributions of the three population to the mean relative velocity. First, and as expected, the agreement between the measured unconditional mean bubble relative velocity as evaluated from equation \ref{eq11} and as directly measured happens to be as good as that on $V_G$. In the homogeneous regime, the respective contributions of the three populations to the mean relative velocity remain the same as for the mean bubble velocity. In the heterogeneous regime, these proportions are changed with 57\% coming from clusters, 42\% arising from intermediate regions and less than 1\% for voids. The contribution of clusters to the relative velocity is thus significantly enhanced (with an increase by more than 10\%) compared with their contribution to the mean bubble velocity.   

For both regimes, the fact that the unconditional bubble velocity is well recovered from bubble velocities conditioned by the three meso-scale structures that were identified indicates that the data processing is reliable. The important point for the dynamics lies in the respective contributions of the three populations. Let us first underline that the results presented above are provided for clusters containing at least one bubble. For both regimes, the figures remain very close when considering a minimum of two bubbles in clusters, so that our analysis is not sensitive to the precise cluster definition. Also, and as already said, the decomposition into meso-scale structures in the homogeneous regime may appear as somewhat artificial, but, owing to their definition, such structures can indeed be identified from a RPP even though their probability of occurrence is low. Let us also recall that the definition of frontiers between populations derives from the specific shape of the Vorono\"i cell pdf in the heterogeneous regime. These beginnings being established, the key point here is that the velocities conditioned by structures happen to be quite different in the two regimes. In particular, the contribution of clusters to both absolute velocities and relative velocities of bubbles is significantly larger in the heterogeneous regime than in the homogeneous regime. These features are clear indications that the physics at play are different, with collective effects present in the heterogeneous regime while the repartition of bubbles and their dynamics remain quasi uniform in the homogeneous regime. 

The role of meso-scale structures on the bubble motion being clarified, it would be worthwhile to develop a prediction of the velocity of bubbles pertaining to each populations. This is the objective of the next section.

\section{Scaling of conditional relative velocities and meso-scale structures dynamics \label{sec5}} 

In \cite{Mezui2022}, we considered an inertia-buoyancy equilibrium at the scale of the bubble column from which we derived the scaling of transport velocities for liquid and gas phases. An equilibrium involving inertia and buoyancy is now assumed at the scale of each meso-scale structure immersed in the two-phase mixture to evaluate the velocity of that meso-scale structure $U_s$ relative to the mean flow of the mixture $U_m$. We borrow here an argument developed by Cholemari \& Arakeri (\cite{cholemari2009axially}) for turbulent flows driven by buoyancy: these authors argue that the velocity $(gL \delta \rho / \rho)^{1/2}$ corresponds to the `free fall' velocity that a coherent region of density $\rho + \delta \rho$ sinking (or creaming) in a medium of density $\rho$ reaches after a distance $L$, and $L$ is such that the flow becomes uncorrelated at distances of order $L$. That `fall velocity' corresponds to $U_s - U_m$. For a meso-scale structure, $\delta \rho$ is the difference in density between the structure and its surroundings. 

In bubble columns, the density of meso-scale structures is $(1-\varepsilon_{structure})\rho_L$ where $\varepsilon_{structure}$ denotes the void fraction averaged at the scale of the meso-scale structure. Meanwhile, the mean density of the two-phase mixture is $(1-\varepsilon)\rho_L$. Hence, $\delta \rho / \rho_L = (\varepsilon-\varepsilon_{structure})$, meaning that the difference in density is proportional to the excess or to the deficit of void fraction in the structure compared with the mean void fraction $\varepsilon$ in the surrounding medium. Typically, clusters - that are gas dominated regions - would have an upward directed (i.e. positive) relative velocity with respect to the mean flow of the gas-liquid mixture in the order of: 

\begin{equation}\label{eq8}
U_{R~cluster-mixture} = U_{cluster}-U_m = C_{cluster} (g L_{cluster} [{\varepsilon_{cluster}-\varepsilon}])^{1/2},
\end{equation}

\noindent where the prefactor $C_{cluster}$ is (a priori) of order one, while voids, that are liquid dominated regions, would have a downward directed (i.e. negative) relative velocity with respect to the mean flow of the gas-liquid mixture:  

\begin{equation}\label{eq9}
U_{R~void-mixture} = U_{void}-U_m = -C_{void} (g L_{void} [{\varepsilon-\varepsilon_{void}}])^{1/2},
\end{equation}

\noindent where $C_{void}$ is a prefactor of order unity. The same reasoning can be applied to intermediate regions, so that the magnitude of the velocity between intermediate regions and the mixture obeys:

\begin{equation}\label{eq10}
U_{R~int-mixture} = U_{int}-U_m= C_{int} (g L_{int} [{\varepsilon_{int}-\varepsilon}])^{1/2},
\end{equation}

\noindent again with a prefactor $C_{int}$ of order unity. A positive sign has been retained for equation \ref{eq10} because, according to Figure \ref{fig19}, the mean void fraction in intermediate regions $\varepsilon_{int}$ is slightly larger than the local void fraction $\varepsilon$ when in the heterogeneous regime (the opposite holds in the homogeneous regime). 

Equations \ref{eq8} to \ref{eq10} connect the relative velocity between a meso-scale structure and the mixture with the meso-scale structure characteristics in terms of size and concentration. The relevance of the propositions \ref{eq8} to \ref{eq10} is tested in the next section. For that, all the necessary information for evaluating the quantities $L_s [\varepsilon_s - \varepsilon] $is available from experiments (see Section \ref{sec3}). We also need to connect the relative velocity $U_s - U_m$ between a meso-scale structure and the mixture with the conditional relative bubble velocity $V_{b \vert s} - V_L$ which is a quantity directly accessible to measurements. 

\subsection{Test of the relevance of the scaling proposed for the meso-scale structure relative velocity}

\begin{figure}
     \centering
     \begin{subfigure}[b]{0.49\textwidth}
         \centering
         \includegraphics[width=\textwidth]{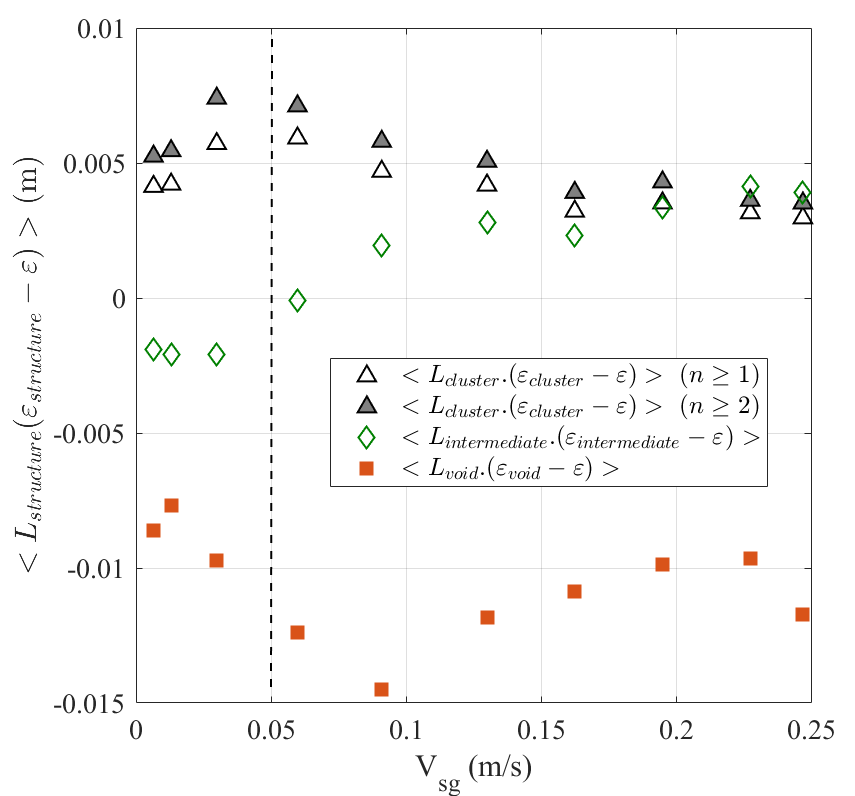}
         \caption{}
         \label{fig20a}
     \end{subfigure}
     \hfill
         \caption{Evolution of the products $\langle L_{cluster} [\varepsilon_{cluster}-\varepsilon]\rangle$ for $n \geq 1$ and for $n \geq 2$, $\langle L_{int} [\varepsilon_{int}-\varepsilon]\rangle$ and $\langle L_{void} [\varepsilon_{void}-\varepsilon]\rangle$  with $V_{sg}$. Measurements performed in a $D=0.4$m column, on the column axis at $H/D=3.625$.}
        \label{fig20}
\end{figure}

Concerning the quantities $L_s [\varepsilon - \varepsilon]$, and as all the data exploited here have been collected on the column axis, the local void fraction $\varepsilon$ for the mixture is equal to $\varepsilon_{axis}$. The mean values $\langle L_{void} [\varepsilon_{void}-\varepsilon]\rangle$, $\langle L_{cluster} [\varepsilon_{cluster}-\varepsilon]\rangle$ and $\langle L_{int} [\varepsilon_{int}-\varepsilon]\rangle$ are shown versus $V_{sg}$ in Figure \ref{fig20}. All these quantities happen to remain fairly stable at large $V_{sg}$, say for $V_{sg}$ above $\approx 10-15$cm/s. Note also that setting the minimum number of bubbles in clusters to 1 or to 2 does not induce any significant difference on $\langle L_{cluster} [\varepsilon_{cluster} -\varepsilon] \rangle$. 

The last ingredient needed to test equations \ref{eq8} to \ref{eq10}, is a connection between the mixture velocity  $U_m$ and the mean liquid velocity $V_L$. By definition, $U_{m}$ is the mixture volumetric flux, that is the velocity of the center of volume of both phases (\cite{ishii1975thermo}). Therefore, $U_{m}$ is related to unconditional phasic velocities by $U_{m} = (1-\varepsilon)U_L+\varepsilon V_G$, and $U_{m}-V_L$ writes

\begin{equation}\label{eq12}
U_{m}-V_L= \varepsilon (V_G-V_L) =\varepsilon U_R.
\end{equation} 

\noindent The prefactors $C_{structure}$ in equations \ref{eq8} to \ref{eq10} can now be evaluated. Indeed, for each meso-scale structure, one has:

\begin{equation}\label{eq13}
U_{R ~ structure-mixture}= U_{s}-U_{m} =U_{s}+V_L-V_L-U_{m}  =V_{b \vert s}-V_L-\varepsilon U_R.
\end{equation} 

\begin{figure}
\centering
\includegraphics[width=\textwidth]{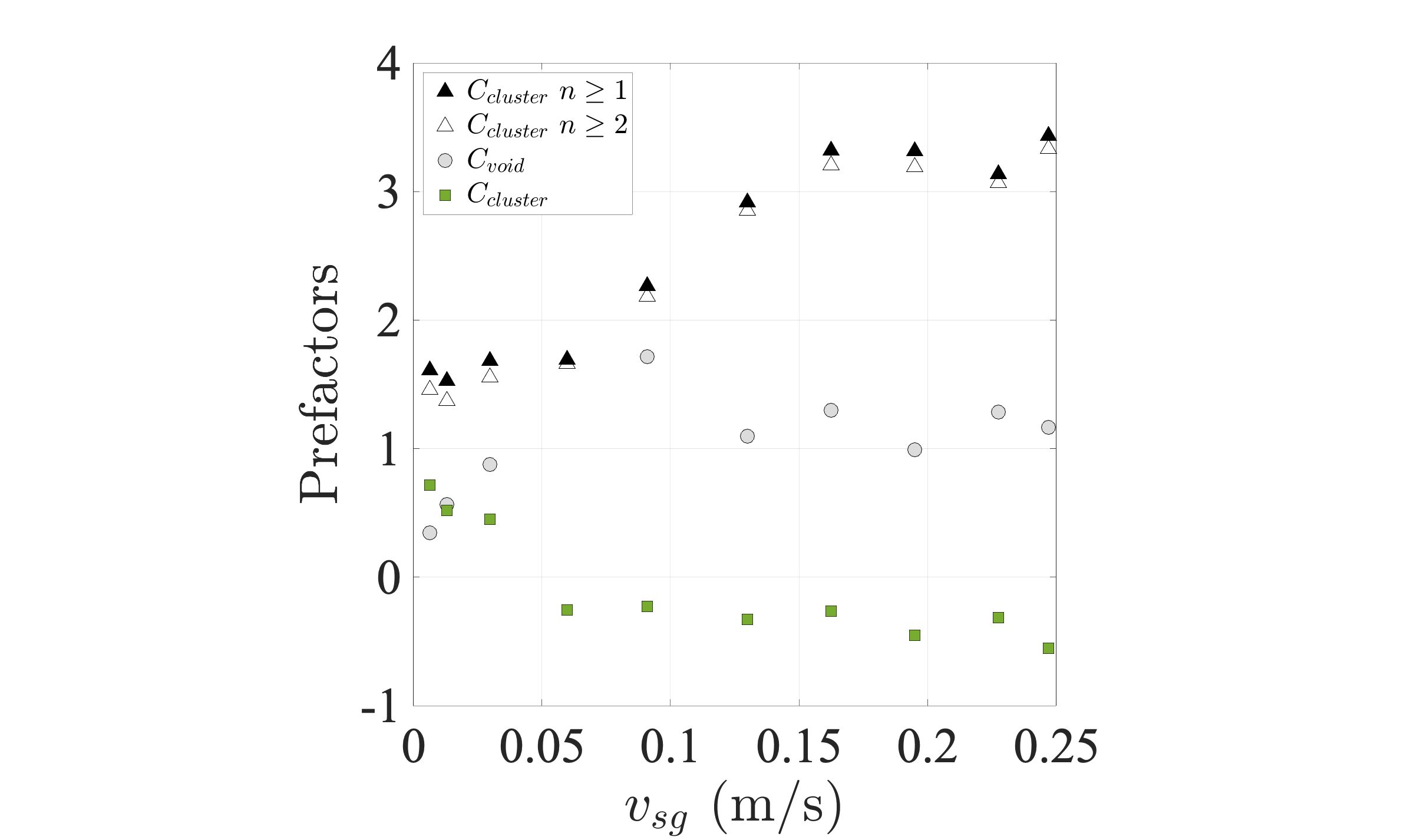}
\caption{Evolution of the prefactors $C_{cluster}$, $C_{int}$ and $C_{void}$ deduced form equation \ref{eq14} combined with equations \ref{eq8} to \ref{eq10} with $V_{sg}$.} \label{fig24}
\end{figure}

\noindent Note that, in the last equality of equation \ref{eq13}, $U_{s}$ has been identified with $V_{b \vert s}$. Strictly speaking, these two quantities are not the same as $V_{b \vert s}$ represents the mean velocity of bubbles within the structure considered while, as discussed in the introduction of Section \ref{sec5}, $U_{s}$ corresponds to the velocity of the whole coherent region forming the structure meaning that $U_{s}$ includes information on both gas and liquid phases. In clusters, owing to their large void fraction in the heterogeneous regime (see Figure \ref{fig19}c), it is reasonable to assume that both phases move at nearly the same velocity and hence that $U_{clusters} \sim V_{b \vert clusters}$. A similar argument could be put forward for intermediate regions when $V_{sg}$ is large. For void regions, experience shows that $V_{b \vert voids}$ and $V_L$ nearly coincide when in the heterogeneous regime (see Figure \ref{fig22}). Hence, assuming $U_{s} \sim V_{b \vert s}$ seems reasonable (to confirm that, liquid velocity measurements conditioned by the local gas concentration would be useful), and equation \ref{eq13} combined with equations \ref{eq8}, \ref{eq9} or \ref{eq10} provides an estimate of the prefactor $C_{s}$, namely:  

\begin{equation}\label{eq14}
C_{s}= U_{R ~ structure-mixture} / (g L_{s} \vert \varepsilon_{s} - \varepsilon \vert)^{1/2},
\end{equation} 

\noindent where the denominator is known from figure \ref{fig20}. The values $C_{s}$ deduced from equation \ref{eq14} are given as a function of $V_{sg}$ in Figure \ref{fig24}. They all tend to nearly constant values at large $V_{sg}$: the mean values of  $C_{s}$ in the heterogeneous regime are provided in Table \ref{tab6}; all these figures are almost insensitive to the range of $V_{sg}$ selected to compute the average.

Moreover, all prefactors are of order unity, with $C_{cluster} \sim 3$, $C_{int} \sim1.2$ and $C_{void} \sim -0.36$. The scalings proposed in equations \ref{eq8} to \ref{eq10} are therefore consistent, and these models provide the correct magnitude of the relative velocity of meso-scale structures with respect to the mixture. These results also confirm that the dynamics of these meso-scale structures is indeed controlled by a buoyancy-inertia equilibrium applied at their respective scales.

\begin{table}
\centering
\resizebox{\textwidth}{!}{%
\begin{tabular}{|l|c|c|c|c|}
\hline
\multicolumn{1}{|c|}{\multirow{2}{*}{\begin{tabular}[c]{@{}c@{}}Measured $U_{R~structure-mixture}/$\\ $(g L_{structure} [\varepsilon_{structure} - \varepsilon])^{1/2}$\end{tabular}}} & $C_{cluster}$ & $C_{cluster}$ & \multirow{2}{*}{\begin{tabular}[c]{@{}c@{}}$C_{int}$ \\ Intermediate regions, eq.\ref{eq10}\end{tabular}} & $C_{void}$ \\
\multicolumn{1}{|c|}{} & Dense   regions $n\geq$1, eq.\ref{eq8} & Dense regions $n\geq$2, eq.\ref{eq8} &  & Void regions, eq.\ref{eq9} \\ \hline
Average prefactor for  $V_{sg}\geq$ 6 cm/s & 2.87 & 2.78 & 1.26 & -0.34 \\ \hline
Average prefactor for $V_{sg}\geq$ 9 cm/s & 3.06 & 2.97 & 1.26 & -0.36 \\ \hline
Average prefactor for $V_{sg}\geq$ 13 cm/s & 3.22 & 3.13 & 1.17 & -0.38 \\ \hline
\end{tabular}%
}
\caption{Mean values of prefactors $C_{s}$ in the heterogeneous regime when assuming $U_{s} = V_{b \vert s}$. }
\label{tab6}
\end{table}

Here, each conditional bubble relative velocity $V_{b \vert s} - V_L$ has been connected with the characteristics of the corresponding meso-scale structure. Hence, thanks to equation \ref{eqN1}, the unconditional bubble relative velocity is related with the characteristics of all three meso-scale structures present in the flow, combined with the repartition of bubbles between these three populations. That result demonstrates that the enhancement of the bubble relative velocity observed in the heterogeneous regime is the direct consequence of the collective dynamics occurring in these buoyancy driven bubbly flows when the gas fraction is large enough. 

\subsection{Scaling of the mean relative velocity: discussion}

At this stage, it is worthwhile to come back to the scaling of the relative velocity of equation \ref{eq1N} that involves the reference velocity $(g D \varepsilon)^{1/2}$ identified in \cite{Mezui2022}. Let first us recall that equation \ref{eq1N} arises from scaling laws for mean velocities of both phases that have been corroborated over a large range of column diameters and flow conditions. However, direct measurements of the relative velocity in the heterogeneous regime such as those presented in Section \ref{sec2} are absent from the literature, so that the dependency of the relative velocity with the column diameter predicted by equation \ref{eq1N} cannot be directly tested from available data. Similarly, the analysis in terms of meso-scale structures developed above concerns but a single column diameter, and more experiments are required to further investigate meso-scale structures characteristics. Yet, as shown in  \cite{Mezui2022}, the existence of an asymptotic heterogeneous state at large $V_{sg}$ is supported by many experiments performed in various column diameters and flow conditions. The presence of such an asymptotic behavior prompted us to rescale the relative velocities $U_{R~structure-mixture}$  using $D$ as the relevant scale for the dimensions $L_s$ of meso-scale structures and using the gas hold-up $\varepsilon$ for scaling the differences in concentration $ [ \varepsilon_{s} - \varepsilon ]$. Table \ref{tab7} provides $L_s/D$,  $[\varepsilon_{s} - \varepsilon]/\varepsilon$ and the quantity $\langle g L_{s} [ \varepsilon_{s} - \varepsilon ] \rangle^{1/2} / (g D \varepsilon)^{1/2}$ that enter equation \ref{eq8} to equation \ref{eq10}. The last line of Table \ref{tab7} provides for each structure the relative velocities $U_R$ structure-mixture scaled by the velocity $(gD\varepsilon)^{1/2}$: the resulting coefficients are comprised between 0.1 and 0.6: such $O(1)$ values seem reasonable.

\begin{table}
\centering
\resizebox{\textwidth}{!}{%
\begin{tabular}{|l|c|c|c|c|}
\hline
 & Clusters & Clusters &  &  \\
\multirow{-2}{*}{Mean values evaluated for $V_{sg}\geq$ 9cm/s} & with n$\geq$1 & with n$\geq$2 & \multirow{-2}{*}{Intermediate regions} & \multirow{-2}{*}{Void regions} \\ \hline
Mean   size & 0.018 m & 0.022 m & 0.062 m & 0.074 m \\ \hline
Mean   size / D & 0.045 & 0.054 & 0.155 & 0.185 \\ \hline
$\langle (\varepsilon_{structure}- \varepsilon) / \varepsilon\rangle$ & 0.464 & 0.585 & 0.054 & -0.42 \\ \hline
$\langle g L_{structure} [\varepsilon_{structure} - \varepsilon ]\rangle^{1/2 }/ (gD\varepsilon)^{1/2}$ & 0.17 & 0.19 & 0.15 & -0.3 \\ \hline
$C_{structure}$ (from Table \ref{tab6}) & 3.06 & 2.97 & 1.26 & -0.36 \\ \hline
prefactor $U_{R~structure-mixture} / (gD\varepsilon)^{1/2}$ & \cellcolor[HTML]{FFFFFF}0.520 & \cellcolor[HTML]{FFFFFF}0.564 & \cellcolor[HTML]{FFFFFF}0.189 & \cellcolor[HTML]{FFFFFF}-0.108 \\ \hline
\end{tabular}%
}
\caption{Average characteristics of meso-scale structures in the heterogeneous regime measured on the axis of a $D=0.4$m bubble column and at $H/D=3.625$, pre-factors $C_{structure}$ et ratio $U_{R~structure-mixture} / (gD\varepsilon)^{1/2}$}
\label{tab7}
\end{table}

The scaling for the unconditional relative velocity $U_R=V_G-V_L$ can be deduced from the above information. Starting from equation \ref{eq11} and still assuming that $U_{s} =V_{b \vert s}$, we have $U_R = N_{clusters} (U_{clusters} - U_{m}) + N_{int} (U_{int} - U_{m}) + N_{voids} (U_{voids} - U_{m}) + \varepsilon U_R$ that transforms into :

\begin{equation}\label{eq15}
\begin{split}
(1-\varepsilon) U_R 
=   N_{clusters} ( U_{clusters} - U_{m}) + N_{int} (U_{int} - U_{m}) \\ 
+  N_{voids} (U_{voids} - U_{m}) =  \large( N_{clusters} \left[(U_{clusters} - U_{m}) / (gD\varepsilon)^{1/2} \right] \\
+ N_{int} \left[(U_{int} - U_{m}) / (gD\varepsilon)^{1/2} \right]  \\
+ N_{voids} \left[(U_{voids} - U_{m}) / (gD\varepsilon)^{1/2} \right]  \large) (gD\varepsilon)^{1/2} 
= C_R (g D \varepsilon)^{1/2}
\end{split}
\end{equation} 

The prefactor $C_R$ has been evaluated over various ranges of $V_{sg}$ within the heterogeneous regime: it is given in Table \ref{tab8} where we have also considered the two options for clusters (namely $n \geq 1$ and $n \geq 2$). Overall, the dispersion is small, and one gets: 

\begin{equation}\label{eq16}
(1-\varepsilon) U_R \sim 0.30 \pm 0.01 (g D \varepsilon)^{1/2}.
\end{equation} 

\noindent As the void fraction in the heterogeneous regime ranges from 20 to 37\% for the experimental conditions considered here, the ratio $U_R/(gD\varepsilon)^{1/2}$ evolves from 0.37 to 0.50, to be compared with the value $U_R \sim 0.41 (gD \varepsilon)^{1/2} $ deduced from direct velocity measurements (see equation \ref{eq1N}). The difference between these two results remains in the interval $[-11\%; +23\%]$. Such a difference is quite acceptable owing to the variety of independent measurements involved in that analysis (the latter include void fraction, unconditional and conditional relative velocities, statistics on size and on concentration for the three meso-scale structures, repartition of bubbles among these structures based on Vorono\"i tessellations) and possibly also owing to the assumption $U_{s} \approx V_{b \vert s}$ we made.

\begin{table}
\centering
\resizebox{\textwidth}{!}{%
\begin{tabular}{|l|c|c|c|c|c|c|}
\hline
\multicolumn{1}{|c|}{$U_{R~structure-mixture}$/} & Cluster & Cluster & Intermediate & Void & $(1-\varepsilon)/(gD\varepsilon)^{1/2}$ & $(1-\varepsilon)/(gD\varepsilon)^{1/2}$ \\
\multicolumn{1}{|c|}{$(gD\varepsilon)^{1/2}$} & n$\geq1$ & n$\geq2$ & regions & regions & n$\geq1$ & n$\geq2$ \\ \hline
mean value   over the range  $V_{sg}\geq$ 6cm/s & 0.517 & 0.556 & 0.176 & -0.105 & 0.292 & 0.307 \\ \hline
mean value   over the range  $V_{sg}\geq$ 9cm/s & 0.520 & 0.564 & 0.189 & -0.108 & 0.300 & 0.317 \\ \hline
mean value   over the range  $V_{sg}\geq$ 13cm/s & 0.515 & 0.563 & 0.187 & -0.110 & 0.297 & 0.315 \\ \hline
\end{tabular}%
}
\caption{Prefactors for the relative velocities $U_{R}$ structure-mixture and for the unconditional relative velocity $U_R$ with respect to the velocity scale $(gD\varepsilon)^{1/2}$ in the heterogeneous regime (from measurements on the axis of a D=0.4m bubble column at H/D=3.625).}
\label{tab8}
\end{table}

The expression of the relative velocity from equation \ref{eq1N} derived from scaling considerations is recovered here using the partition of bubbles into three distinct populations. Again, and as seen in Section \ref{sec4}, equation \ref{eq15} and equation \ref{eq16} show that the increase of the bubble relative velocity beyond the terminal velocity value originates from the dynamics of the meso-scale structures present in the heterogeneous regime. In this process, clusters bring the strongest contribution. Thanks to the significant proportion of bubbles they gather (Table \ref{tab3}) and thanks to their high relative velocity with respect to the mixture (clusters correspond to the largest ratio $U_{R~structure-mixture}/(gD\varepsilon)^{1/2}$ in Table \ref{tab7}), they contribute by 65-68\% to $U_R$. Intermediate regions host the majority of bubbles but their relative velocity with respect to the mixture is about 3 times smaller than that of clusters: they contribute by 33-35\% to $U_R$. Last, void regions are sinking in the mixture: they carry few bubbles and their (negative) contribution to $U_R$ is almost negligible ($\sim$ 2\%). 

Although data are lacking to directly examine how the relative velocity evolves with the column diameter, equation \ref{eq15} and \ref{eq16} provide an indirect way to discuss the dependency of $U_R$ on $D$. According to experiments, the void fraction weakly varies with $D$ (in \cite{Mezui2022}, the void fraction on the column axis is found to evolve as $D^{-0.2}$), so that equation \ref{eq16} indicates that the relative velocity monotonously increases with $D$. At first sight, that prediction seems odd if one refers to a single bubble dynamics that is controlled by its interaction with the liquid at a scale commensurable with the bubble size, and not with the size of the domain. However, we have seen that collective dynamics in these buoyancy driven bubbly flows plays a central role in the formation of meso-scale structures, and that the presence of both dense and dilute structures drives the momentum exchange between phases and leads to an enhancement of the relative velocity. Further, the various contributions $U_{R~ structure-mixture}$ appearing in equation \ref{eq15} are directly related with the characteristics in terms of size and concentration that each meso-scale structure has in the asymptotic limit (i.e. in the limit of large $V_{sg}$). The question left is thus wether the scalings with $D$ and with $\varepsilon$ used to built Table \ref{tab7} are relevant or if they are artificial. 

Concerning $\varepsilon$, an examination of the evolution of the coefficients $C_s$ of Figure \ref{fig24} versus $\varepsilon$ instead of $V_{sg}$ indicates that a scaling with the void fraction is indeed acceptable for gas hold-up above about 30\%. 

Regarding the dimensions of meso-scale structures, it is unlikely that the typical width of clusters grows with $D$, but the void regions do have an extension of order $D$. Such a statement is supported by our experiments in the $D=0.4$m column (Figure \ref{fig19}). It is also sustained by the results presented in the next section where it is shown that void regions correspond to large-scale vorticity regions those dimension is of order $D$.

\subsection{Flow structure in the heterogeneous regime}

\begin{figure}
\centering
\includegraphics[width=0.6\textwidth]{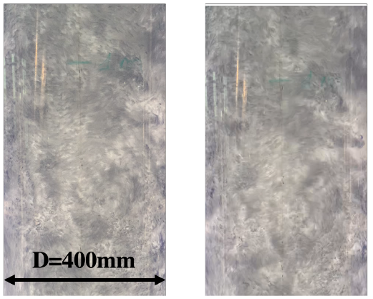}
\caption{Images of flow near column walls in the heterogeneous regime. Bright zones indicate the presence of many bubbles while dark zones correspond to structures comprising few bubbles. The vertical extent of the images corresponds to sections at 0.8 and at 2 meters above gas injection ($D=0.4$m, static liquid height = 2m). Movies are included in the supplementary material of \cite{Mezui2022}.} \label{figCol}
\end{figure}

Concerning the internal structure of the flow in the heterogeneous regime, the spatial organisation of the gas phase deduced from 1D Vorono\"i tessellations has demonstrated the presence of thin regions at high void fraction, and of large regions at low void fraction. When examining the flow through column walls using direct lightning, these void regions correspond to dark zones (Figure \ref{figCol}) while bright regions indicate a significant presence of bubbles (those interface reflects light back toward the observer). These dark regions seem to correspond to the large-scale vortical-like structures that have often been reported in the literature (and that are illustrated in the video enclosed as supplementary material associated to the reference \cite{Mezui2022}).  

To quantify such vortical structures, we exploited the local liquid velocity provided by a Pavlov tube. Spatial correlations were not accessible with a single sensor, and liquid velocity measurements conditioned by the local gas concentration were not attempted. Instead, we considered time series collected from a single Pavlov tube even though its temporal resolution was limited (about 14Hz). 

The zero crossings of the signal $v_L(t) - V_L$ were detected and the density of zero crossing $n_s$ per unit length was evaluated using the mean liquid velocity to transform time into space. The resulting characteristic length scale $n_s^{-1}$ provides the mean vertical size of vortical structures. Let us mention that, for a turbulent single-phase flow, Liepmann \& Robinson (\cite{liepmann1953counting}) related the Taylor microscale $\lambda$ to the average distance $n_s^{-1}$ between zero crossings of a streamwise velocity signal of a turbulent flow, and they show that $n_s^{-1}=B\lambda$ where $B$ is a constant that accounts for intermittency ($B= \pi$ for a Gaussian time series with also a Gaussian derivative). In our case, it is not clear if the vertical spatial scale we construct is related to any turbulence scale, as we are far from the conditions of homogeneity and isotropy required by the model from Liepmann \& Robinson. From measurements in the $D=0.4$m column, $n_s^{-1}$ was found equal to D (within 10\%) for $V_{sg}$ larger than $\approx 10$ cm/s. Although acquired in a single column, this result supports the idea that the size of vortical structures does scale as the bubble column diameter. A few other results on the integral length scale of turbulence (\cite{Mezui2023}) and on the size of vortical structures in various bubble columns (\cite{cartellier2019bubble}) also indicate that $D$ is the relevant scale. 

In parallel, let us show that the regions containing bubbles are not in the form of `compact' clusters of bubbles. Indeed, the high particle Reynolds number bubbles considered here (see section \ref{sec2}) are in a constant drag coefficient regime (when isolated). If one considers a compact, close to spherical assembly of $N$ such bubbles, the dynamics of that ensemble would be also governed by a constant drag coefficient, and its relative velocity would be equal to $N^{1/6} U_T$. Therefore, the relative velocities between about $2 U_T$ and $2.5 U_T$ that we measured in heterogeneous conditions at large $V_{sg}$ (for $V_{sg}$ above $10$cm/s, as discussed in Section \ref{sec2}), would be recovered with $N=64$ for $2 U_T$, or with $N=244$ for $2.5 U_T$: these figures are $10$ to $50$ times larger than the average number of bubbles detected in clusters (Figure \ref{fig19}). Clearly, the existence of compact assemblies of bubbles does not correspond to observations. The question is now how thin bubbly `sheets' could induce such an enhancement of the bubble relative velocity.

\begin{figure}
\centering
\includegraphics[width=0.5\textwidth]{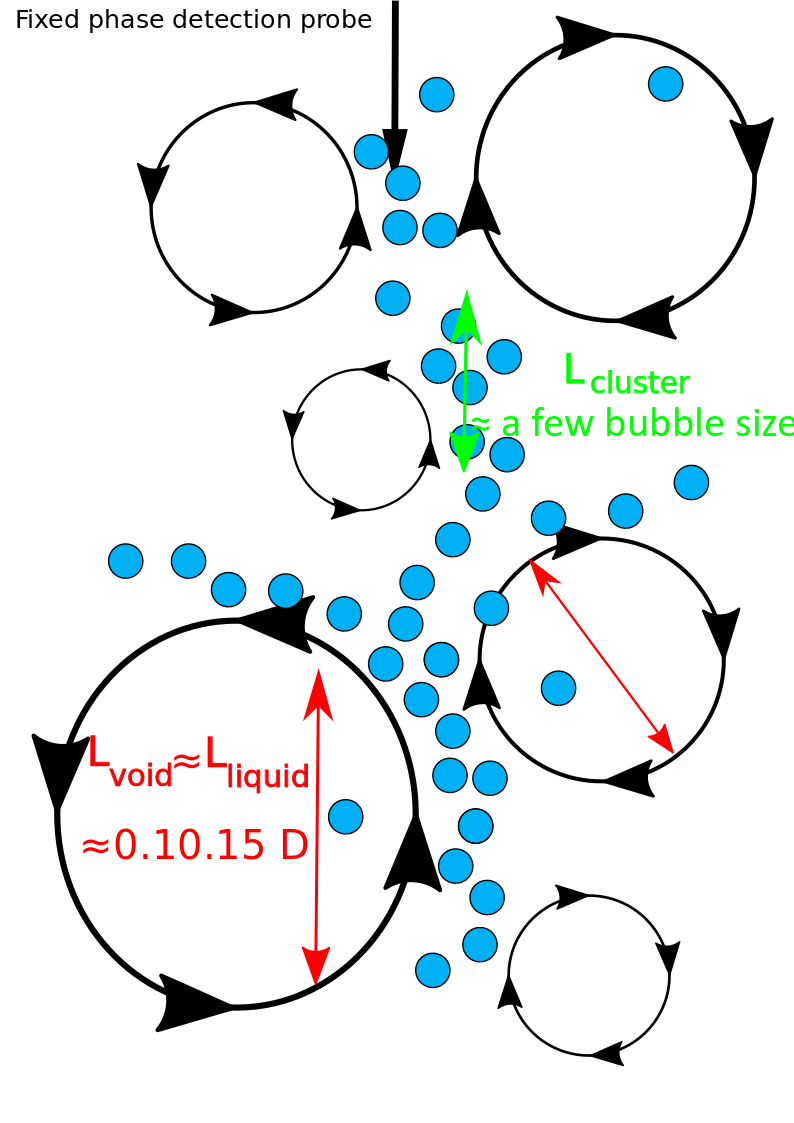}
\caption{Tentative sketch of the flow organization.} \label{fig29}
\end{figure}

From these findings, a tentative cartoon of the spatial organisation of phases in the heterogeneous regime emerges: bubbles accumulate in narrow (a few bubbles in size) regions located in between large (typically $\approx 0.2 D$ according to Figure \ref{fig19}b) vortical structures that are almost free of bubbles. This situation is sketched in Figure \ref{fig29}. The dense regions containing most bubbles are like thin `sheets' or `curtains' (a few bubble diameter wide) located in between vortices whose extent is of order $D$. Thus, these arguments combined with equation \ref{eq15} support a relative velocity controlled by the lateral dimension of the column. 

To conclude, in this section the relative velocity observed in the heterogeneous regime has been connected with the characteristics in terms of size and concentration of dense, intermediate and dilute regions formed in these buoyancy driven bubbly flows. These meso-scale structures and the dynamics they induce are believed to be at the origin of the swarm factor introduced in Eulerian two-fluid simulations to evaluate the momentum exchange between phases in the heterogeneous regime. Moreover, a number of experimental results support the existence of an asymptotic flow organisation at large $V_{sg}$, in particular, with the saturation of the gas concentration in dense regions, and with limiting values of the spatial extent of meso-scale structures. Additional investigations are required to fully determine how these asymptotic values evolve with parameters, and in particular with the column diameter. Let us finally underline that, in their simulations of heterogeneous conditions, \cite{panicker2020computational} captured the presence of bubble swarms with a characteristic length scale of order $V_G^2/g$ and predicted a significant increase of the mean gas velocity compared with homogeneous conditions: these findings are consistent with the experimental results presented here. 

\section{Velocity fluctuations, internal structure and fast-track mechanism \label{sec6}}

So far, we have discussed the scalings of the mean transport velocity and of relative velocities. Let us consider now velocity fluctuations. \cite{Mezui2022}, found that the unconditional standard deviations of liquid velocity $V_L'$ and of bubble velocity $V_G'$ evolve as $(gD\varepsilon)^{1/2}$, but, these results were mostly validated for a single bubble column diameter. 

\begin{figure}
\centering
\includegraphics[width=\textwidth]{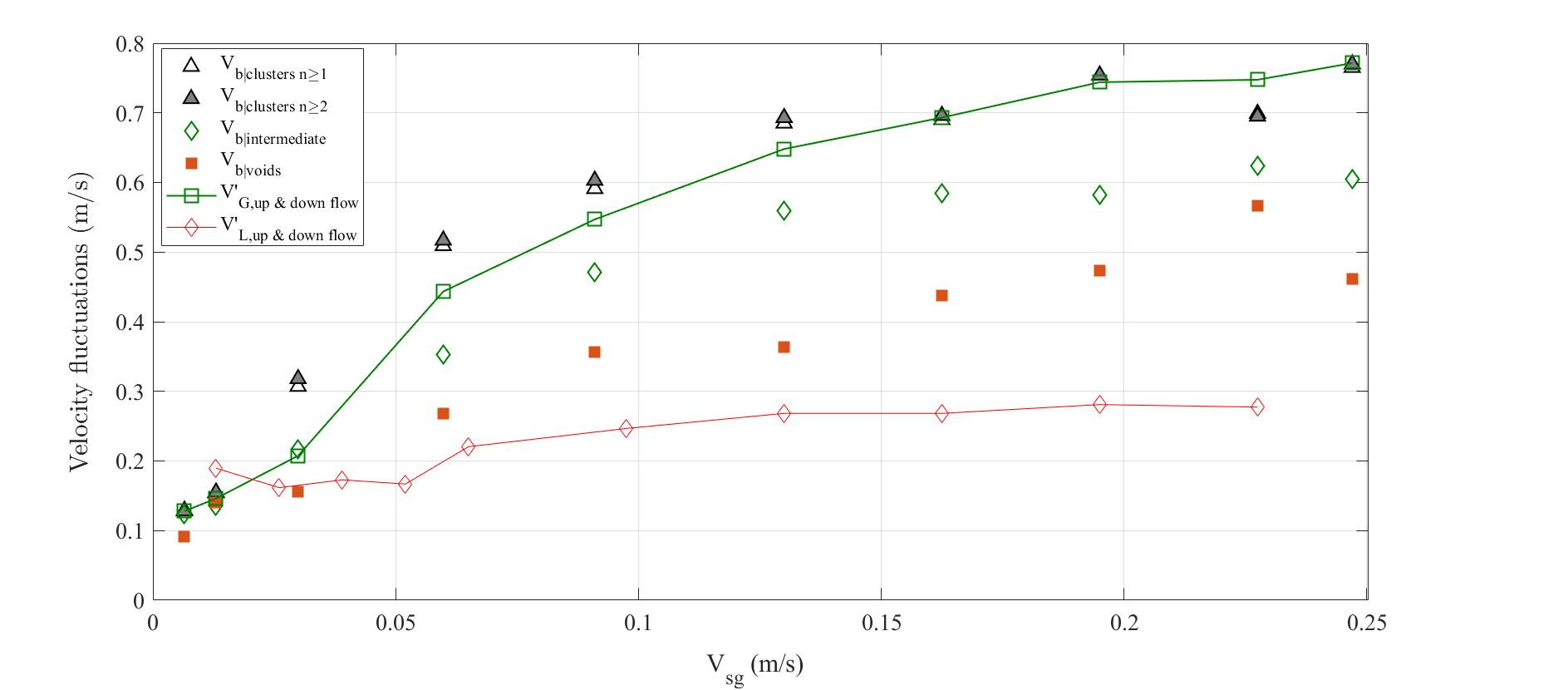}
\caption{Standard deviations $std(v_{b \vert s} - V_G)$ of bubble velocity conditioned by meso-scale structures i.e. clusters, intermediate and void regions with respect to the unconditional mean bubble velocity $V_G$ versus the gas superficial velocity. Comparison with unconditional standard deviations of liquid and gas velocities. Measurements performed in a $D=0.4$m column, on the column axis at $H/D=3.625$.} \label{fig25}
\end{figure}

Before discussing the possible origin of such a scaling, let us analyse the experimental data gathered on the axis of the D=0.4m column. As before, we take advantage of the Doppler probe to examine the behavior of the standard deviation of the bubble velocity conditional to meso-scale structures. The mean unconditional bubble velocity $V_G$ as the reference, and we evaluated the standard deviation $V'_{b \vert s} = std(v_{b \vert s} - V_G)$ where std denotes the standard deviation, and where $v_{b \vert s}$ is the instantaneous bubble velocity in the selected structure and where std denotes the standard deviation, are shown in Figure \ref{fig25} for clusters, intermediate regions and void regions. Their evolutions of the quantities $v_{b \vert s}$ with $V_{sg}$ have qualitatively the same allure as those of mean conditional velocities $V_{b \vert s}$ shown in Figure \ref{fig22}. Notably, the differences between meso-scale structures remain small when in the homogeneous regime and increase when in the heterogeneous regime. In the latter regime, the difference from one structure to the other typically amount to $0.1 - 0.2$m/s. Also, velocity fluctuations are larger in clusters than in intermediate regions, and are larger in intermediate regions than in void regions. These trends are not sensitive to the minimum number of bubbles in a cluster.

As for the mean velocity, the unconditional standard deviation of bubble velocity can be deduced from the contributions of the three meso-scale structures, weighted by the proportion of bubbles they contain. Indeed, the sum $N_{clusters} V'_{b \vert clusters} + N_{int} V'_{b \vert int}  + N_{voids} V'_{b \vert voids}$ has been compared to the standard deviation $V_G'$, and the agreement is good with a discrepancy of at most 25\% in the homogeneous regime and at most 20\% in the heterogeneous regime. In the latter case, the contributions to velocity fluctuation mainly originate from clusters (42-45\% contribution) and from intermediate regions (51-55\% contribution) with a small remaining contribution (3-4\%) arising from void regions. 

The fact that bubble velocity fluctuations are significantly larger than liquid velocity fluctuations in the heterogeneous regime has already been reported (\cite{Mezui2022}).  We show here that this is also true for conditional bubble velocities, with the exception of the very low $V_{sg}$ limit that belongs to the homogeneous regime. This is not surprising owing to the overwhelming contributions of clusters and of intermediate regions to bubble velocity fluctuations in the heterogeneous regime. Oddly, this is even true in void regions, possibly because these statistics recover bubbles having quite diverse environments as the local void fraction typically ranges from 0.4 to 0.1 times the mean hold-up (see Figure \ref{fig15} and associated comments).  

\begin{figure}
\centering
\includegraphics[width=\textwidth]{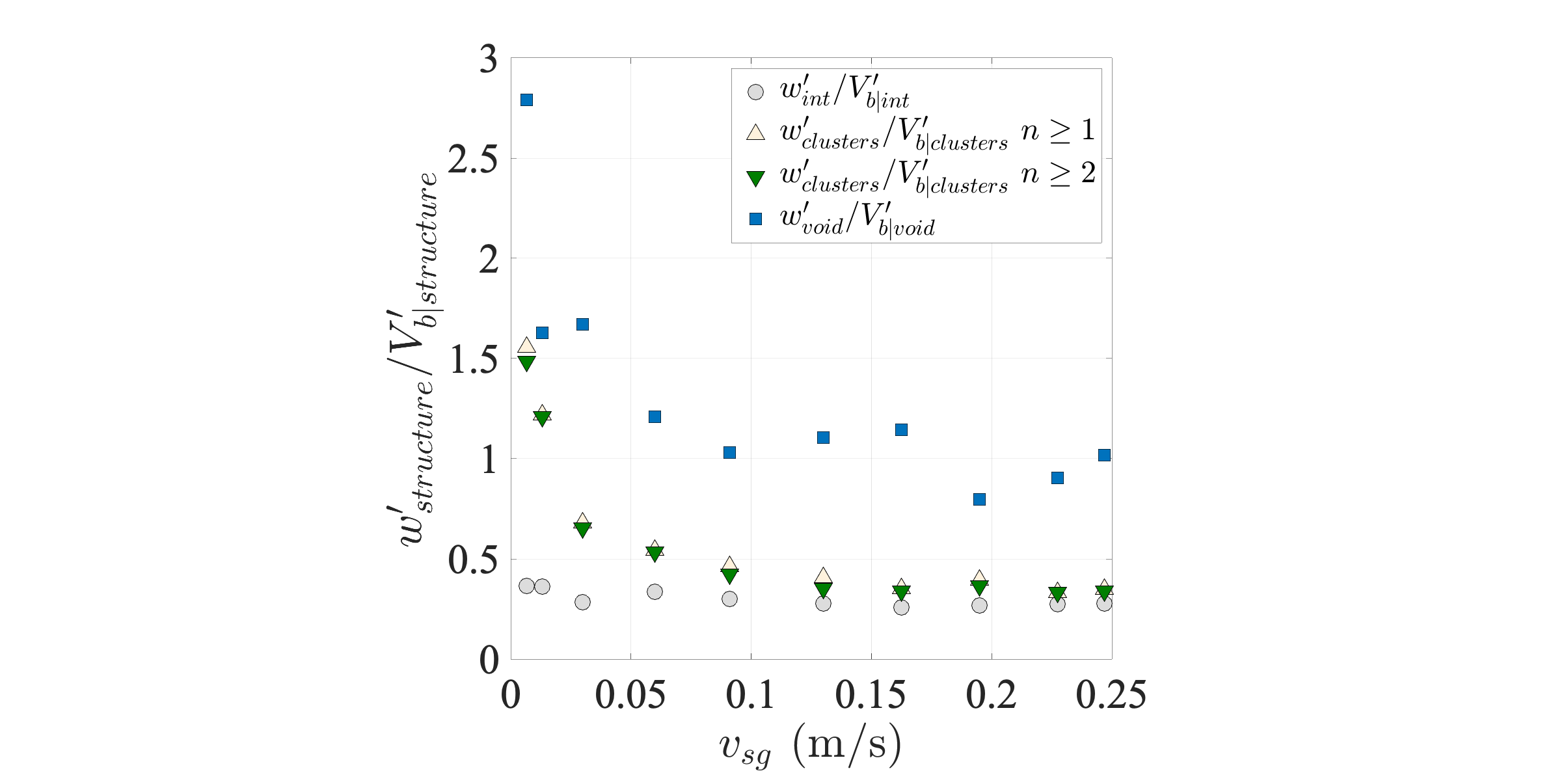}
\caption{Comparison of the velocity variation $w'_{structure}$ due to size and concentration of meso-scale structures pertaining to the same population with bubble velocity fluctuations conditioned by meso-scale structures $V'_{b \vert structure}$. Measurements performed in a $D=0.4$m column, on the column axis at $H/D=3.625$.} \label{fig26}
\end{figure}

The fluctuations in bubble velocity arise from bubble velocity variations between different types of meso-scale structures. They can also arise from velocity variations between meso-scale structures belonging to the same population as both buoyancy and inertia are variable from one structure to the other. In an attempt to evaluate that second contribution, we considered the quantities $L_{structure} [\varepsilon_{structure} - \varepsilon]$ that enter equations \ref{eq8} to \ref{eq10} for the three meso-scale structures. Following the scaling rules given by equations \ref{eq8} to \ref{eq10}, the velocity fluctuation associated with variations in size and concentration for a given meso-scale structure is evaluated as $w'_{structure} = g std(L_{structure} [\varepsilon_{structure} - \varepsilon])^{1/2}$. These estimations are compared with the bubble velocity fluctuations conditioned by meso-scale structures in Figure \ref{fig26}. The curves are rather stable for $V_{sg}$ above 10-15cm/s. It happens that $ w'_{structure} / V'_{b \vert structure} \sim 3$ for clusters, 1.5 for intermediate regions and about 1 for void regions. The contribution of the variability in size and in concentration of structures is therefore small for clusters, and it remains moderate but still higher than unity for intermediate regions. Hence, as clusters and intermediate regions bring the largest contribution to fluctuations, a significant fraction of bubble velocity fluctuations is therefore related with the velocity differences between the various types of meso-scale structures. Owing to the key contribution of the latter, we develop hereafter the idea that a fast track mecanisms is at play and that it connects the relative velocity with bubble velocity fluctuations. 

According to the presumed internal organisation of the flow in the heterogeneous regime as sketched in Figure \ref{fig29}, a plausible mechanism could be the following. As bubbles are mainly located between vortices, a fast track mechanism similar to the one observed in turbulent flow laden with inert particles (\cite{wang1993settling}) takes place. Bubbles are channeling between vortices, and they preferentially pick up the side of eddies with an upward motion (choosing the downward side induces a much larger local relative velocity and hence a much larger drag). The neat result is a faster upward directed vertical bubble velocity. Such a picture is consistent with the conditional velocity measurements presented above. It also provides a physical background to the swarm coefficient often introduced in simulations to force the drag on a bubble to decrease with the local void fraction. 

Such a picture is also consistent with the impact a fast track mechanism has on the enhancement of the relative velocity. Indeed, in turbulent flows laden with inert particles, the enhancement of the settling velocity of dense particles is found proportional to the velocity fluctuations of the background turbulence, with a prefactor typically between 0.1 and 0.5 depending on particle and on flow characteristics (see for e.g. \cite{wang1993settling,mora2021effect}). Making a crude parallel with the present situation, we can consider $V'_L$ as the magnitude of external velocity fluctuations. In section \ref{sec3}, we have seen that, on the axis of a D=0.4m column, the liquid fluctuations scale as $V'_L \approx 0.22 (gD\varepsilon)^{1/2}$ while the mean relative velocity scales as $U_R \approx 0.3-0.4 (gD\varepsilon)^{1/2}$: these two results indicate that, when in the heterogeneous regime, $U_R$ and $V'_L$ remain proportional with a ratio about $0.5-0.7$. This finding is therefore consistent with what is known about the impact of a fast-track mechanism on the relative motion of inclusions with respect to a turbulent continuous phase. Beside, the velocity enhancement observed here is consistent with recent results concerning the frontier between enhancement and hindering (\cite{mora2021effect}):  the Rouse number of inclusions $Ro=U_R/V'_L$, that lies here between 0.5 and 0.7, is indeed small enough to avoid the triggering of a loitering scenario. 

A question left open at this stage is why bubbles remain (in average) accumulated and stuck in between large-scale liquid structures. This is a counter-intuitive organization if one thinks of bubbles interacting with turbulent eddies in a denser fluid, as bubbles are preferentially moving towards low pressure zones, i.e. in the core of eddies. Our belief is that the situation in bubble columns is not the same as that of bubbles immersed in a weak turbulent field. We have shown in section \ref{sec2} that, in the heterogeneous regime, buoyancy is the source of the mean motion and of velocity fluctuations by way of internal density gradients. In such flows, the accumulation or the depletion of bubbles are not governed by eddies interacting with independent, quasi-isolated bubbles, but by collective dynamics that imposes its forcing on the more inert phase. In other words, a local bubble accumulation induces an upwelling motion that must be compensated by a nearby downward motion of an essentially liquid (possibly including a few bubbles) zone. This is why, once formed, the (thin) clusters of bubbles as well as the empty regions are believed to persist for some time which is long enough compared with the transit time of the mixture from the bottom to the top of the bubble column. In our experiments, that transit time is about 1.5 to 3 seconds in the heterogeneous regime, and it is indeed small compared with the bubble response time $a^2/\nu_L$ that is about 10 seconds here. In other words, for the flow conditions considered here, there is not enough time available for bubbles to be significantly dispersed or to significantly diffuse outside dense regions. Note that similar dynamics have been reported in experiments (\cite{kimura1995bubble}) and in simulations (\cite{nakamura2020linear,climent1999large}) of micro-bubbles induced convection in shallow conditions with the formation of mushrooms similar to those arising in Rayleigh-Taylor instability.

The proposed scenario deserves to be tested further using experiments and/or direct numerical simulations. This scenario is also expected to change when considering different flow conditions, notably in terms of coalescence efficiency. As a crude quantification of the limit of validity of this scenario, let us evaluate the size that would have bubbles so that their terminal velocity (when isolated) equals the relative velocity we measured in the $D=0.4$m column. For a relative velocity about $2U_T$ where $U_T \approx 0.21-0.23$m/s, the bubble diameter should be multiplied by $44^{1/3} \approx 3.5$ compared with the size of the bubbles we considered. For a relative velocity about $2.5U_T$, the multiplication factor would be $244^{1/3} \approx 6$. Hence, we expect the proposed scenario to be modified when bubbles above $20$ to $40$mm in equivalent diameter start to appear in the flow. Such a limit also corresponds to bubbles whose terminal velocity, about $0.5-0.7$m/s, would become comparable to the magnitude of velocity fluctuations at large $V_{sg}$ in a $D=0.4$m column. In other words, the Rouse number of such bubbles would become of order one or above, and loitering could occur instead of fast track, and that may possibly lead to a decrease in the relative velocity (\cite{mora2021effect}). Pushing the limit even further, for bubble size of the order of $D$ (as observed for example in fluidised beds), the dynamic would drastically change as one approaches slug flows.


\section{Conclusions}

We pursued the investigation of the hydrodynamics of bubble columns operated in the heterogeneous regime in controlled conditions by focusing on variables conditioned by the local void fraction. Experiments in a $D=0.4$m air-water bubble column were achieved with bubbles pertaining to the wobbling regime and in absence of significant coalescence. 

The concentration field at small scales and its connection with the relative motion between phases has been investigated for gas superficial velocities up to $25$cm/s. From Vorono\"i tessellations in one-dimension built from the signal delivered by an optical probe, the homogeneous/heterogeneous transition has been shown to correspond to a standard deviation of the probability density of Vorono\"i cell width that levels off from its value for a Random Poisson Process (RPP). That departure from a RPP allows to unambiguously identify meso-scale structures, namely clusters (i.e. regions where bubbles tend to accumulate), void regions (i.e. liquid regions including few bubbles) and intermediate regions. These meso-scale structures have been characterised in terms of size and of concentration. In the heterogeneous regime, size and concentration pdfs seem to asymptote as $V_{sg}$ increases. In particular, the mean size of these meso-scale structures tends towards constant values for $V_{sg}$ higher than $\approx 0.1$m/s. Also, the absolute concentration in clusters saturates to 45-50\% while the concentration in voids and in intermediate regions slightly but continuously increases with the mean gas hold-up. The picture that comes out from conditional measurements using optical probes and from Pavlov tubes, comprises void regions that correspond to vortices in the liquid those size is a fraction of $D$, and `thin' clusters -- typically a few bubbles wide -- structured as sheets in between these vortices. This picture is consistent with a fast track mechanism that, for moderate Rouse numbers, leads to liquid velocity fluctuations that are a fraction of the relative velocity between phases.

The origin of the large relative velocity observed in the heterogeneous regime has been for a long time a central question in the hydrodynamics of bubble columns. A series of arguments demonstrating the key role of meso-scale structures on the relative velocity of bubbles has been presented. Notably:

\begin{enumerate}

\item Direct measurements of the unconditional mean relative velocity show that the relative velocity levels off at the homogeneous-heterogeneous transition. Besides, the relative velocity asymptotes at large $V_{sg}$: the limit, in the $D=0.4$m column, is about 2.4 times the terminal velocity of bubbles. 

\item Bubble velocity measurements conditional upon the local gas concentration indicate that bubbles in clusters are moving up must faster, up to 3 to 3.5 times the terminal velocity, than bubbles in void regions those speed is nearly equal to the unconditional liquid velocity. Similarly, bubbles in intermediate regions are moving up faster, up to 1.5 to 2 times the terminal velocity, than bubbles in void regions. 

\end{enumerate}

As the mean unconditional relative velocity of bubbles is recovered from conditional mean relative velocities weighted by the proportion of bubbles present in each meso-scale structure, these findings demonstrate that the flow dynamics in the heterogeneous regime originates from collective effects linked with the apparition of meso-scale structures.

In addition, by assuming equilibrium between inertia and buoyancy at the scale of each meso-scale structure we succeeded to relate the velocity of each meso-scale structure relative to the mixture with its characteristics in terms of size and concentration. Consequently, the unconditionnal mean bubble relative velocity has been related with the characteristics of all three meso-scale structures present in the flow. That result opens the way to the identification of the proper scaling of the relative velocity. In particular, the spatial extension of void regions seems to be proportional to the bubble column diameter, a feature that could explain why the relative velocity evolves as $(gD\varepsilon)^{1/2}$ as indicated by unconditional measurements. These proposals need to be tested over a wider range of conditions. In particular, the impact of bubble column diameters on the relative velocity and meso-scale structures need to be investigated. By examining higher gas superficial velocities, it would be also worthwhile to identify what controls the limits in concentration of meso-scale structures. 


\subsection*{Acknowledgements}
The LEGI is part of the LabEx Tec21 (Investissements d'Avenir  - grant agreement n. ANR-11-LABX-0030). That research was also partially funded by IDEX UGA (n. ANR-15-IDEX-0002) The authors report no conflict of interest.

\appendix

\section{Evaluation of the gas concentration in a Vorono\"i cell}\label{appB}

\subsection{Connection between cell width and cell concentration}

The magnitude of $\Delta T_k$ is related with the local and instantaneous concentration, that is the concentration at the scale of the $k^{th}$ Vorono\"i cell. A large $\Delta T_k$ means that neighbouring bubbles are far from the test bubble, or equivalently that the concentration in the vicinity of the test bubble is low. Inversely, a small $\Delta T_k$ indicates the presence of close neighbours, that is a high void fraction in the vicinity of the test bubble. 

In \cite{raimundo2019hydrodynamics}, it is argued that the quantity $\Delta T_k/\langle \Delta T\rangle$ equals the ratio of the local and instantaneous gas concentration $\varepsilon_k$ to the average gas hold-up $\varepsilon$ at the measuring location, i.e. $\Delta T_k/\langle \Delta T\rangle=\varepsilon/\varepsilon_k$. This equality must be replaced by the equation \ref{eqapp} below. Indeed, let us consider $N$ bubbles detected over a measuring duration $T_{probe}$. By definition, the void fraction $\varepsilon$ equals $(\sum_i t_{gi})/T_{probe}$ where $i$ goes from $1$ to $N$. By construction, the Vorono\"i cells map all the space (that is the whole measuring duration) so that $T_{probe} = \sum_k \Delta T_k$ where $k$ goes from $1$ to $N$. Hence $\varepsilon=(\sum_i t_{gi})/T_{probe} = (\sum_i t_{gi})/ \sum_k \Delta T_k = N \langle t_g\rangle / [N \langle \Delta T\rangle]$ where mean values have been introduced in the last equality. Meanwhile, the local void fraction $\varepsilon_k$ (i.e. at the scale of the $k^{th}$ cell) is $t_{gk} /\Delta T_k$. Therefore :

\begin{equation}\label{eqapp}
\varepsilon_k / \varepsilon = \left[  t_{gk} / \Delta T_k \right] /  \left[ \langle t_g\rangle / \langle \Delta T\rangle \right] =  \left[  t_{gk} / \langle t_g\rangle \right] /  \left[ \langle \Delta T\rangle / \Delta T_k \right].
\end{equation}

\noindent The concentration $\varepsilon_k$ in the $k^{th}$ cell scaled by the local concentration $\varepsilon$ is indeed proportionnal to $\langle \Delta T\rangle / \delta T_k$, but these quantities are not equal. The proportionality coefficient happens to depend on the gas residence time $t_{gk}$ divided by the mean gas residence time $\langle t_g\rangle$. That coefficient changes from one bubble to another in a given record. Therefore, it is not possible to univocally transform a threshold in $\Delta T_k / \langle \Delta T\rangle$ (such a threshold is used to distinguish the three populations, namely clusters, voids and intermediate regions) into a threshold in terms of cell concentration. Instead, the distributions of actual concentrations in each population need to be analysed (see Figure \ref{fig16} and related text). 

\subsection{Gas concentration in a Vorono\"i cell}

\begin{figure}
\centering
\includegraphics[width=\textwidth]{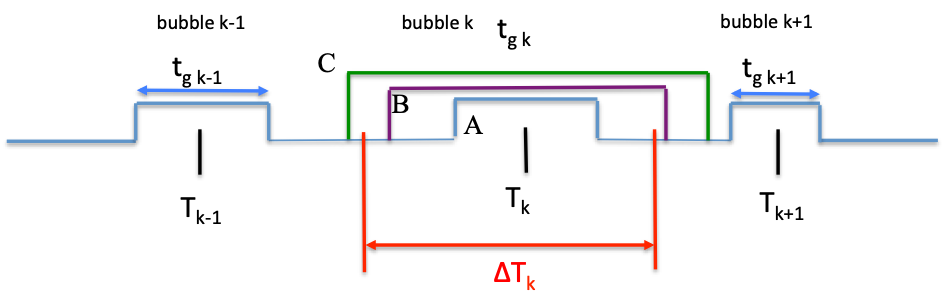}
\caption{Void fraction at the scale of a Vorono\"i cell.} \label{figlast}
\end{figure}

Here above, the void fraction $\varepsilon_k$ at the scale of the $k^{th}$ cell is estimated as $t_{gk} / \Delta T_k$. That formula is exact for the situation of the bubble indicated as A in Figure \ref{figlast}. Let us consider the three successive bubbles $ k-1$, $k$ and $k+1$. Let us increase the residence time of bubble k while maintaining everything else fixed. In particular, the centers of the three bubbles remain located at $T_{k-1}$, $T_k$ and $T_{k+1}$, so that the $k^{th}$ Vorono\"i cell keeps its width $\Delta T_k$. When the $k^{th}$ bubble grows and reaches the situation B (Figure \ref{figlast}), the right hand side of the gas residence interval becomes located outside the cell. When the bubble grows further and reaches the situation C (Figure \ref{figlast}), the residence time of the bubble $k$ exceeds the Vorono\"i cell width and $\varepsilon_k$ becomes larger than unity. Hence, the formula $t_{gk}/ \Delta T_k$ becomes incorrect when dealing with large bubbles (large $t_{gk}$) in a dense surrounding (small $\Delta T_k$). In practice, one should account only for the fraction of the gas residence located inside the Vorono\"i cell. In Section \ref{sec5}, we used a somewhat crude correction as we simply set $\varepsilon_k=1$ whenever $t_{gk}/ \Delta T_k$ exceeded unity. However, the number of events concerned by that issue is quite limited (it is always less than 12\% of the population) so that this approximation does not affect the trends identified nor the mean values. 

\end{document}